\documentclass[a4paper,fleqn,usenatbib]{mnras}

\usepackage[T1]{fontenc}
\usepackage{ae,aecompl}

\usepackage{graphicx}	
\usepackage{amsmath}	
\usepackage{amssymb}	
\usepackage{subfigure}
\usepackage{multirow}
\usepackage{xcolor}

\newcommand{\gcc}{\,{\rm g \, cm}^{-3}}
\newcommand{\pcs}{\,{\rm cm}^{-2}}

\newcommand{\msun}{\, {\rm M}_\odot}

\newcommand{\pc}{\, {\rm pc}}

\newcommand{\ssum}{\displaystyle\sum}

\newcommand{\myr}{\, {\rm Myr}}

\newcommand{\kms}{\, {\rm km \, s^{-1}}}

\newcommand{\subB}{_{_{\rm B}}}

\newcommand{\subFRAG}{_{_{\rm FRAG}}}

\newcommand{\subMAX}{_{_{\rm MAX}}}

\newcommand{\subO}{_{_{\rm O}}}

\newcommand{\subpreS}{_{_{\rm preS}}}

\newcommand{\subSPH}{_{_{\rm SPH}}}

\newcommand{\fwhm}{\mbox{\sc fwhm}_{_\Sigma}}

\newcommand{\rdiv}{W_{_{\!\rm{AS}}}}

\title[Filament widths]{The origin of a universal filament width in molecular clouds}

\author[Priestley \& Whitworth]{
F. D. Priestley\thanks{Email: priestleyf@cardiff.ac.uk} and A. P. Whitworth
\\
School of Physics and Astronomy, Cardiff University, Queen's Buildings, The Parade, Cardiff CF24 3AA, UK \\
}

\date{Accepted XXX. Received YYY; in original form ZZZ}

\pubyear{2021}

\begin{document}
\label{firstpage}
\pagerange{\pageref{firstpage}--\pageref{lastpage}}
\maketitle

\begin{abstract}
   Filamentary structures identified in far-infrared observations of molecular clouds are typically found to have full-widths at half-maximum $\sim\!0.1\pc$. However, the physical explanation for this phenomenon is currently uncertain. We use hydrodynamic simulations of cylindrically-symmetric converging flows to show that the full-width at half-maximum of the resulting filament's surface density profile, $\fwhm$, is closely related to the location of the accretion shock, where the inflow meets the boundary of the filament. For inflow Mach Number, ${\cal M}$, between 1 and 5, filament $\fwhm$s {fall} in the range $0.03\pc\lesssim\fwhm\lesssim0.3\pc$, with higher ${\cal M}$ resulting in narrower filaments. A large sample of filaments, seen at different evolutionary stages and with different values of ${\cal M}$, naturally results in a peaked distribution of $\fwhm$s similar in shape to that obtained from far-infrared observations of molecular clouds. {However, unless the converging flows are limited to ${\cal M} \lesssim 3$, the peak of the distribution of $\fwhm$s is below the observed $\sim 0.1 \pc$.}
\end{abstract}

\section{Introduction}

The internal structure of molecular clouds is highly filamentary \citep{andre2010,andre2014}, and the densest filaments appear to play a key role in star formation. Indeed, the majority of {\it low-mass} stars appear to form in filaments \citep{konyves2015,konyves2020}, and filaments also act to channel material into regions forming {\it high-mass} stars \citep{peretto2013, peretto2014, hacar2018, williams2018, watkins2019, anderson2021}. Understanding how filaments form and evolve is therefore critical to understanding star formation.

A remarkable property of filaments in molecular clouds is that they seem to have a characteristic width of $\sim 0.1 \pc$ \citep{arzoumanian2011,arzoumanian2019}. Although concerns have been raised about the methods used in deriving a `width' from observational data, and about the variance of the underlying distribution \citep{smith2014,panopoulou2017}, this characteristic width does seem to be a general property of filaments, and therefore requires an explanation. Filament widths measured via molecular line emission span a much larger range of values \citep{panopoulou2014,hacar2018,suri2019,alvarez2021,schmiedeke2021}, but these are still consistent with an underlying $\sim\!0.1\pc$ width when chemical and radiative transfer effects are taken into account \citep{priestley2020}.

There is a large body of theoretical and numerical work on the structure and evolution of interstellar and prestellar filaments. Much of this has been based on the assumption of infinitely long, cylindrically symmetric filaments, and has explored -- using variously the Virial Theorem, similarity solutions, semi-analytic formulations and numerical simulations -- their (magneto-)hydrostatic equilibria, their stability against collapse and/or longitudinal fragmentation, and their dynamical formation and evolution. Non-magnetic treatments, assuming an isothermal or polytropic equation of state, include \citet{StodolkiewiczJ1963}, \citet{ostriker1964}, \citet{LarsonRB1985}, \citet{InutsukaSMiyamaS1992}, \citet{KawachiTHanawaT1998}, \citet{HoldenLetal2009}, \citet{FischeraJMartinP2012}, \citet{HeitschF2013a,HeitschF2013b}, \citet{LouY2015} and \citet{TociCGalliD2015a}.  The effects of longitudinal, transverse and helical magnetic fields have been considered (in the ideal limit) by \citet{ChandrasekharFermi1953}, \citet{StodolkiewiczJ1963}, \citet{LarsonRB1985}, \citet{NagasawaM1987}, \citet{FiegeJPudritzR2000a,FiegeJPudritzR2000b}, \citet{HennebelleP2003}, \citet{TilleyDPudritzR2003}, \citet{ShadmehriM2005}, \citet{TomisakaK2014}, \citet{TociCGalliD2015b}, \citet{LouYXingH2016} and \citet{KashiwagiRTomisakaK2021}. In addition, \citet{HansenCetal1976} have explored the stability of spinning infinite filaments against adiabatic perturbations, and \citet{ToalaJetal2012} have evaluated the freefall collapse of infinite filaments.

There is also a body of work that treats filaments of finite length, both with magnetic fields \citep[e.g.][]{SeifriedDWalchS2015}, and without \citep[e.g.][]{BastienP1983, RouleauFBastienP1990, ArcoragiJetal1991, BastienPetal1991, BonnellIBastienP1991, BonnellIBastienP1992, BonnellIetal1991, BonnellIetal1992, BurkertAHartmannL2004, PonAetal2011, PonAetal2012b, ClarkeSWhitworthA2015, ClarkeSetal2016, ClarkeSetal2017}. Much of this work has been concerned with evaluating the competition between different longitudinal fragmentation modes, for example sausage modes versus end-modes, and their potential for forming binary systems.

However, few of these papers take account of how a filament forms in the first place, and the interaction between the evolving structure of a filament and the simultaneous accretion flows onto it. In principle, similarity solutions can take this into account, but in practice the flows derived {\it at early times and on large scales} seem unlikely to represent what occurs in nature. The power of similarity solutions is that more complicated numerical simulations seem often to converge on similarity forms {\it at late times and on small scales}, when gravitational acceleration becomes important; this is true in both cylindrical and spherical symmetry. Moreover, by their nature, similarity solutions do not in general select a particular physical scale.

\citet{kirk2015} and \citet{federrath2016} both find that the typical filament width of $\sim\!0.1\,\pc$ is reproduced in simulations of turbulent, self-gravitating clouds, with or without additions such as magnetic fields. \citet{federrath2016} attributes this to the transition between supersonic and subsonic turbulence, which gives roughly the correct length scale for typical molecular cloud properties. In these cloud-scale simulations, filamentary structures form in regions where the velocity field is convergent  \citep{smith2016,priestley2020}.

In this paper, we consider an idealised model for filament formation, involving a cylindrically-symmetric convergent inflow at Mach Number ${\cal M}$. Our hydrodynamical experiments suggest that for supersonic flows (${\cal M} \ge 1$), a key feature is the accretion shock at radius $\rdiv$ (i.e. distance from the axis of symmetry; subscript `{\sc as}' for {\sc a}ccretion {\sc s}hock) where the convergent flow impacts the central filament. The full-width at half-maximum of the surface-density profile (hereafter the $\fwhm$; subscript `$\Sigma$' for surface-density) is closely related to $\rdiv$, and for moderately supersonic flows we find $\fwhm \sim 2\rdiv {\lesssim\,}0.1 \pc$. A distribution of evolutionary stages and ${\cal M}$ values produces a peaked distribution of $\fwhm$s similar to that observed, although with a somewhat lower peak value unless ${\cal M}$ has a maximum value $\sim 3$.

We use both Cartesian coordinates, $(x,y,z)$, and cylindrical polars, i.e. $w\!=\![x^2+y^2]^{1/2}$, $\phi\!=\!\tan^{-1}(y/x)$, $z=z$. The $z$ axis is the spine of the filament, and so $w$ is radial distance from the spine. In addition, for an observer outside the filament, we define the impact parameter, $b$, which is the projected distance between the line of sight and the $z$ axis.

\begin{figure*}
\centering
\subfigure{\includegraphics[width=\columnwidth]{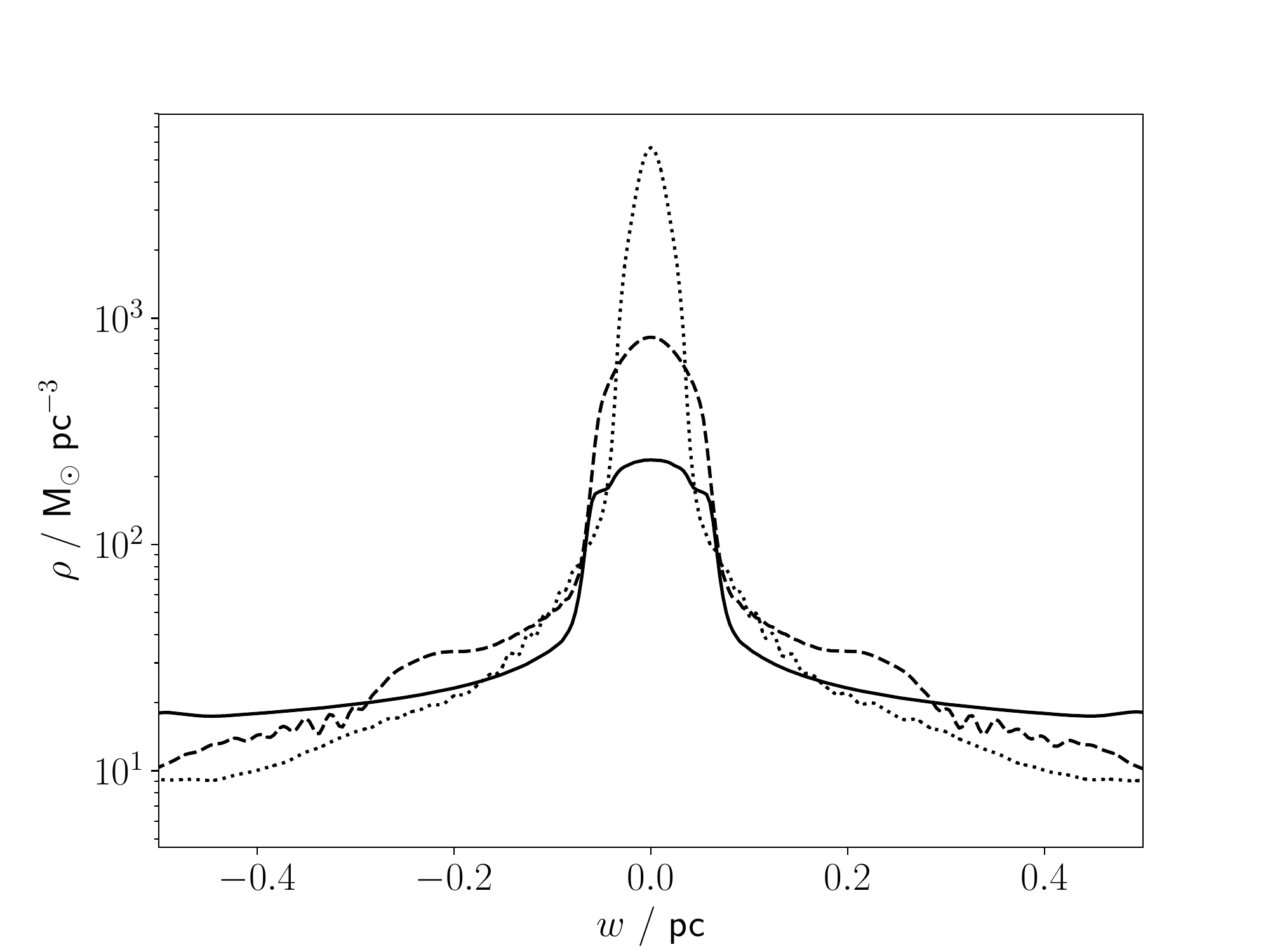}}\quad
\subfigure{\includegraphics[width=\columnwidth]{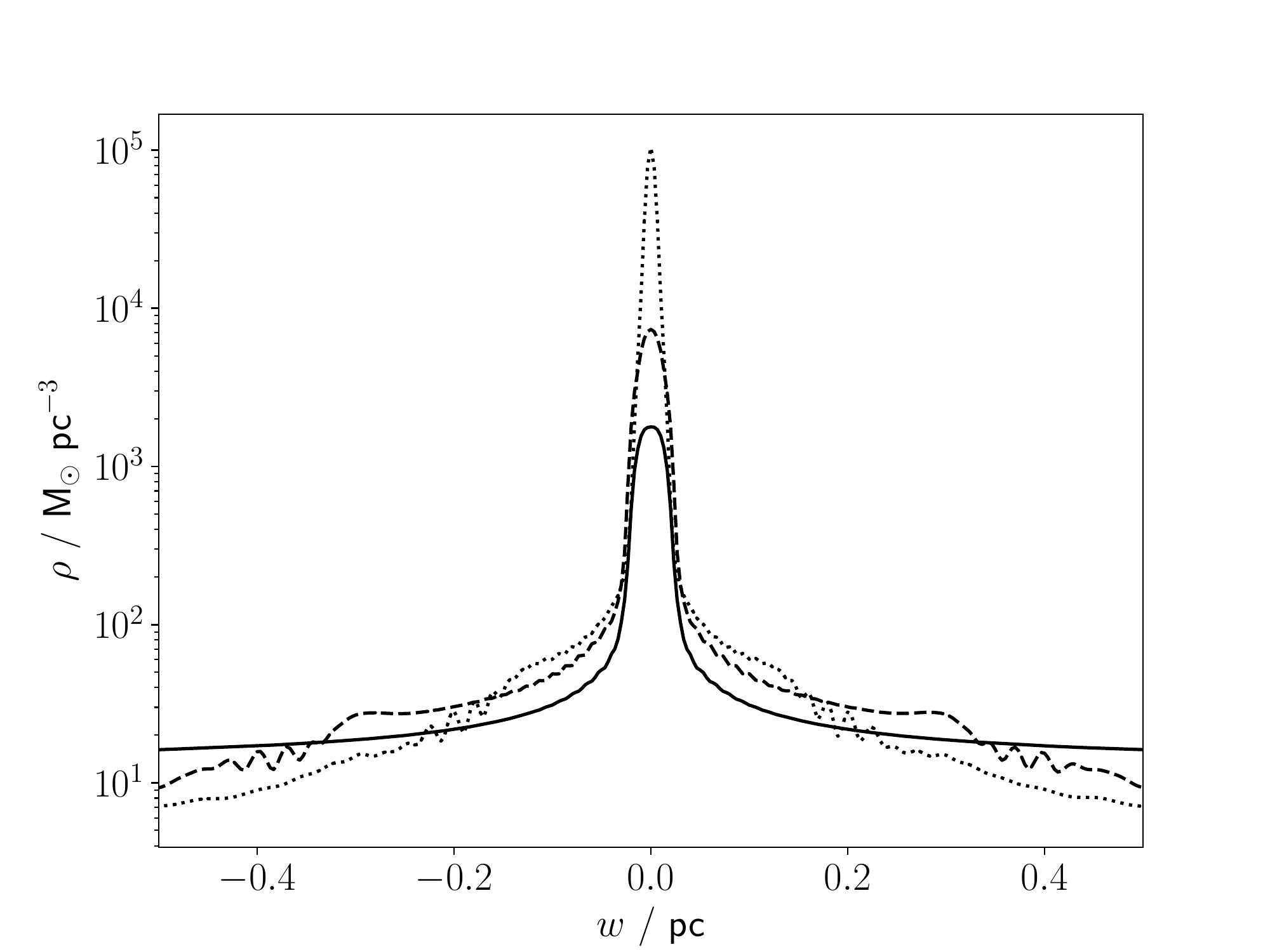}}\\
\caption{{\it Left panel:} volume-density profiles for the model with ${\cal M}=2$ at times $t=$ $0.75\,\myr$ (full line), $1.12\,\myr$ (dashed line), and $1.34\,\myr$ (dotted line; the end of the simulation). {\it Right panel:} volume-density profiles for the model with ${\cal M}=4$ at times $t=$ $0.37\,\myr$ (full), $0.67\,\myr$ (dashed), and $0.82\,\myr$ (dotted; the end of the simulation).}
\label{FIG:VolDensProfs}
\end{figure*}

\begin{figure*}
\centering
\subfigure{\includegraphics[width=\columnwidth]{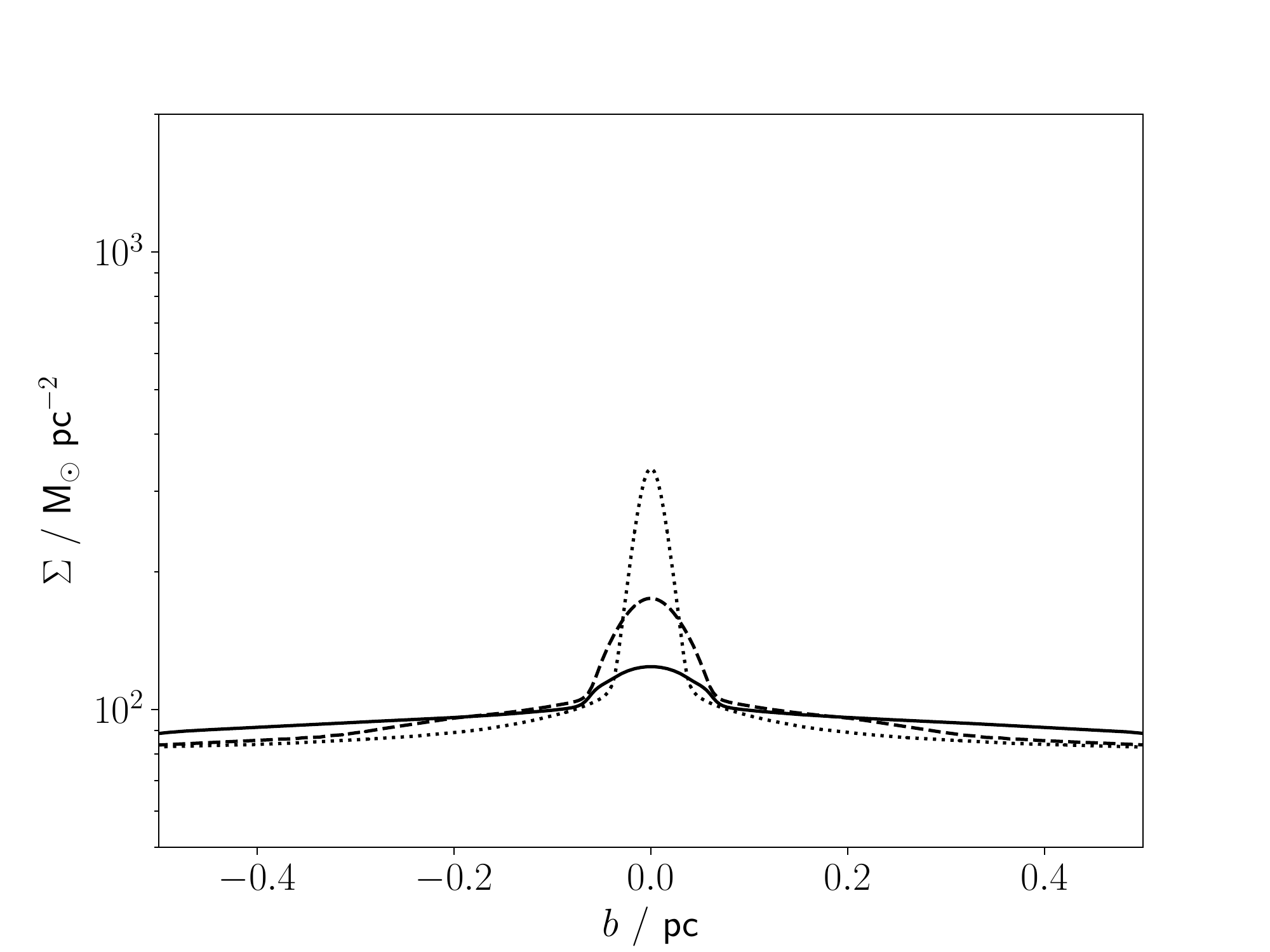}}\quad
\subfigure{\includegraphics[width=\columnwidth]{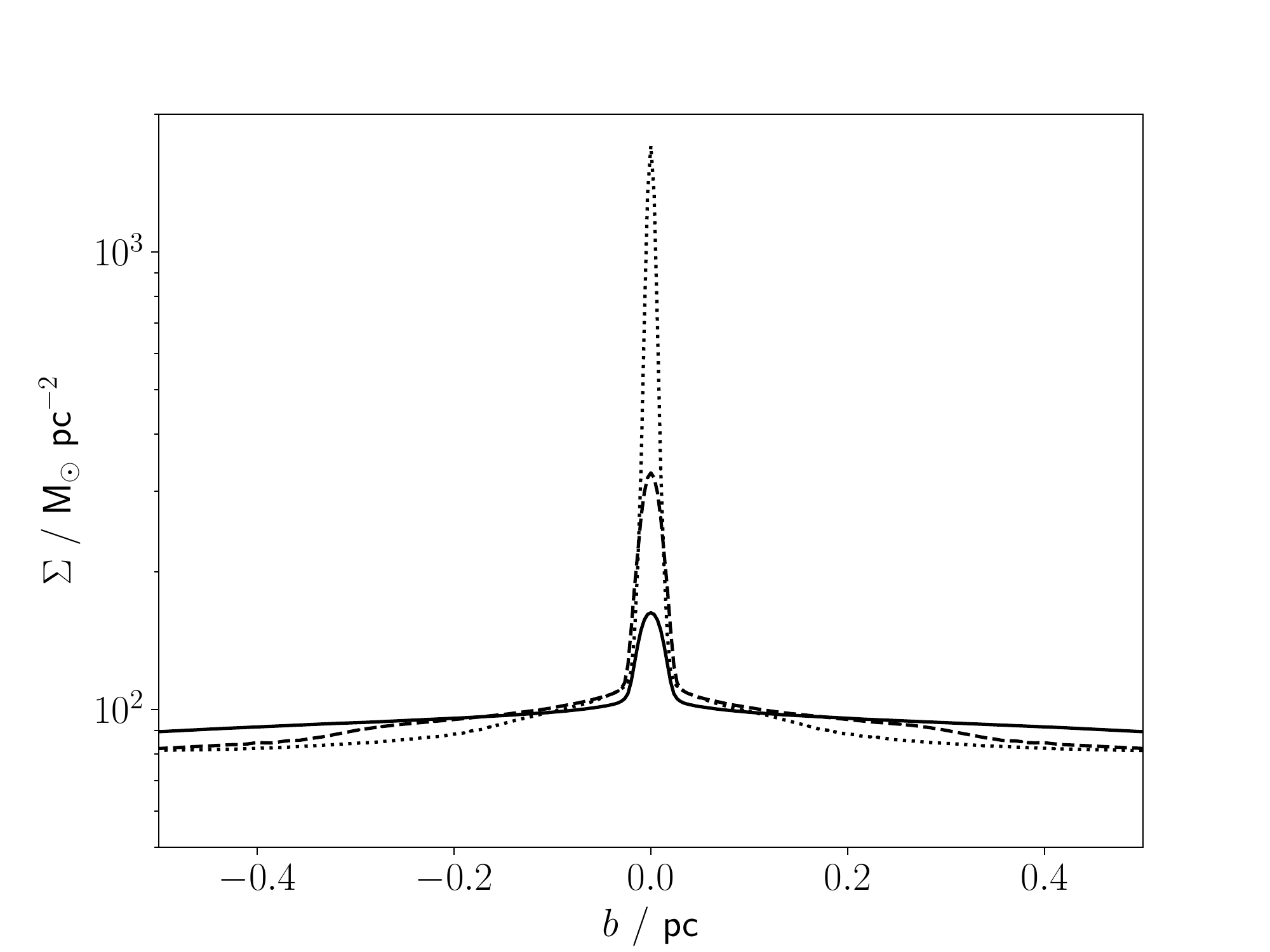}}\\
\caption{As Figure \ref{FIG:VolDensProfs} but showing surface-densities rather than volume-densities.}
\label{FIG:SurfDensProfs}
\end{figure*}

\begin{figure*}
\centering
\subfigure{\includegraphics[width=\columnwidth]{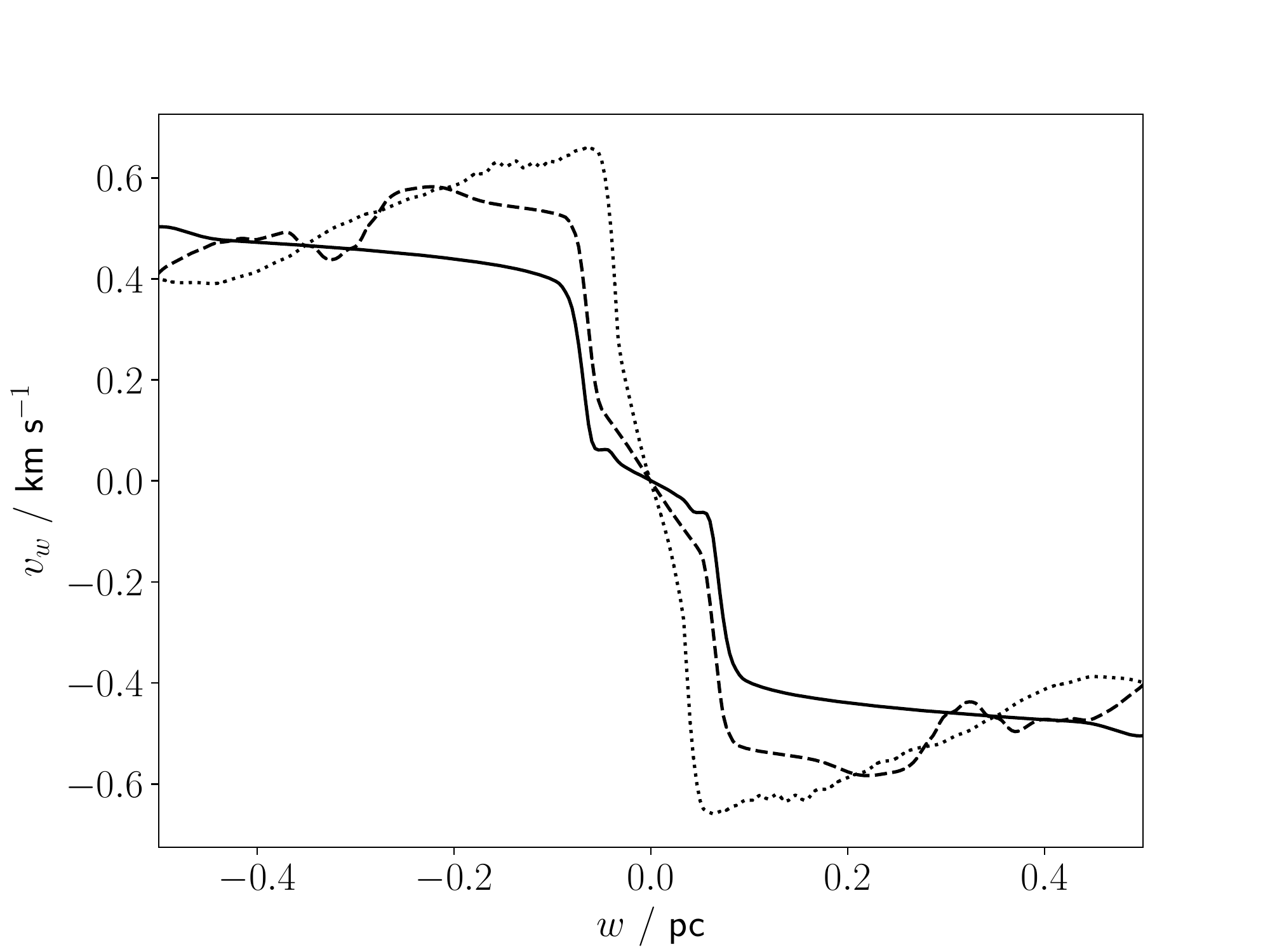}}\quad
\subfigure{\includegraphics[width=\columnwidth]{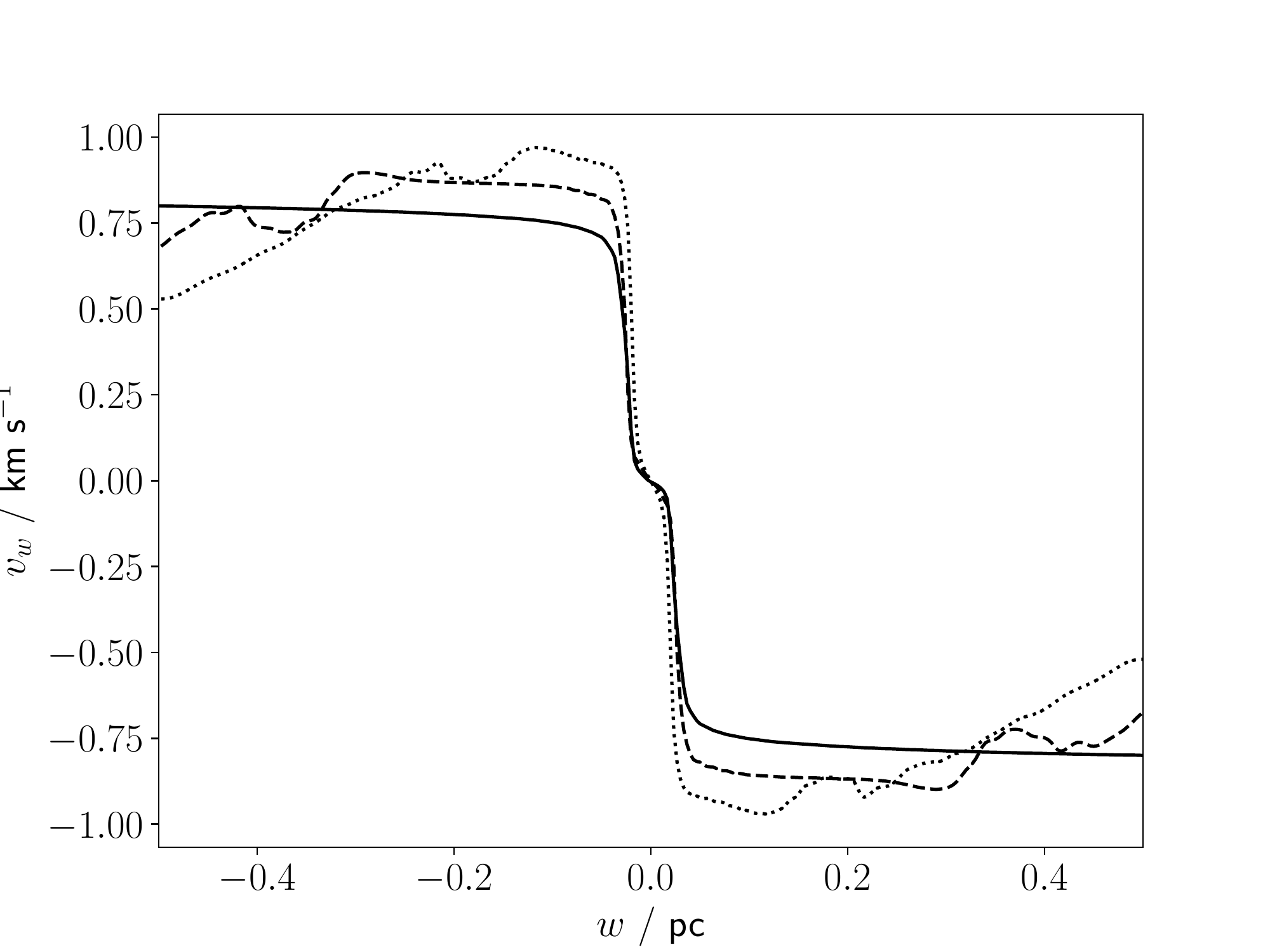}}\\
\caption{Velocity profiles corresponding to the volume-density profiles in Figure \ref{FIG:VolDensProfs}.}
\label{FIG:VelProfs}
\end{figure*}

\begin{figure*}
\centering
\subfigure{\includegraphics[width=\columnwidth]{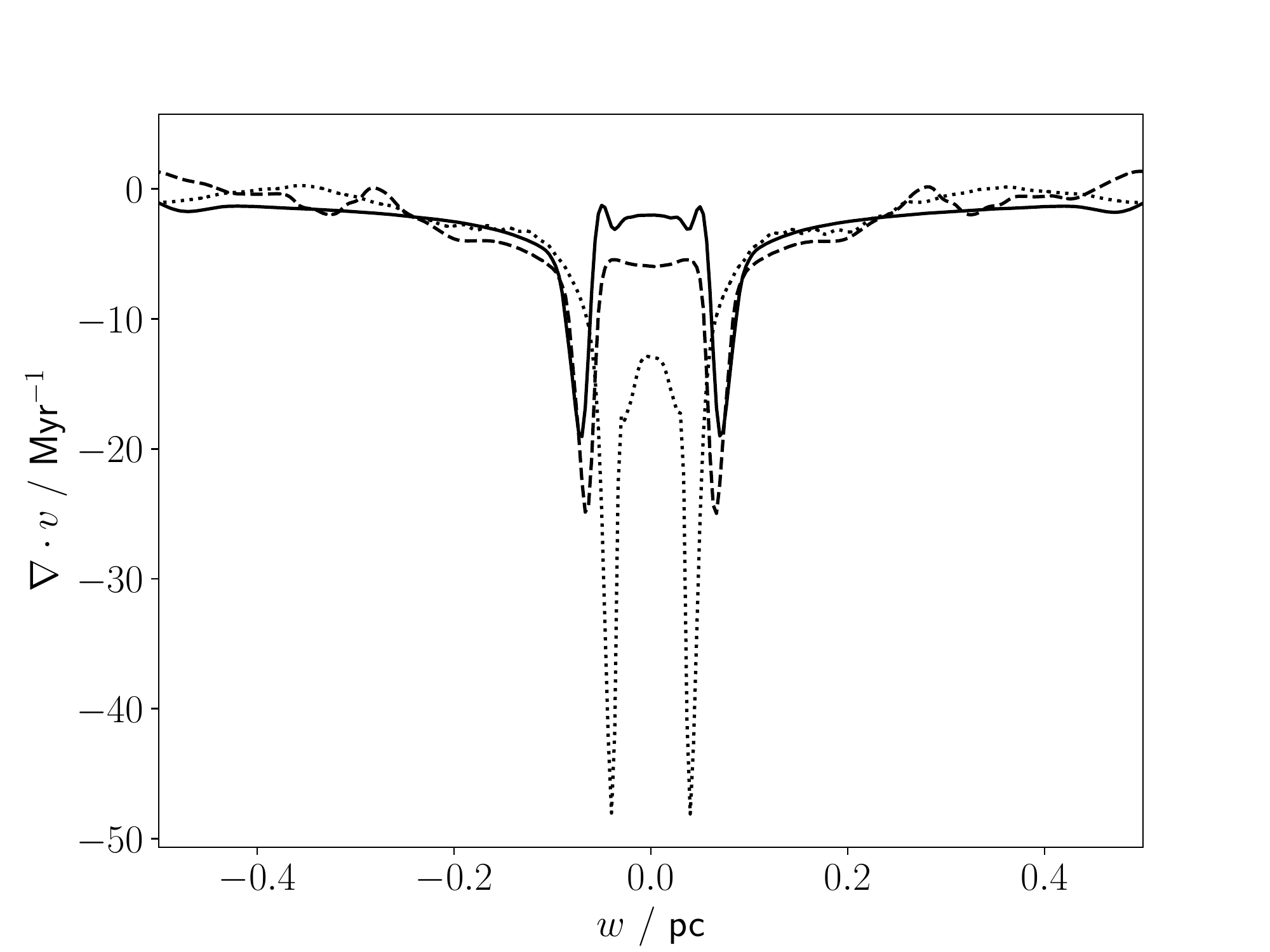}}\quad
\subfigure{\includegraphics[width=\columnwidth]{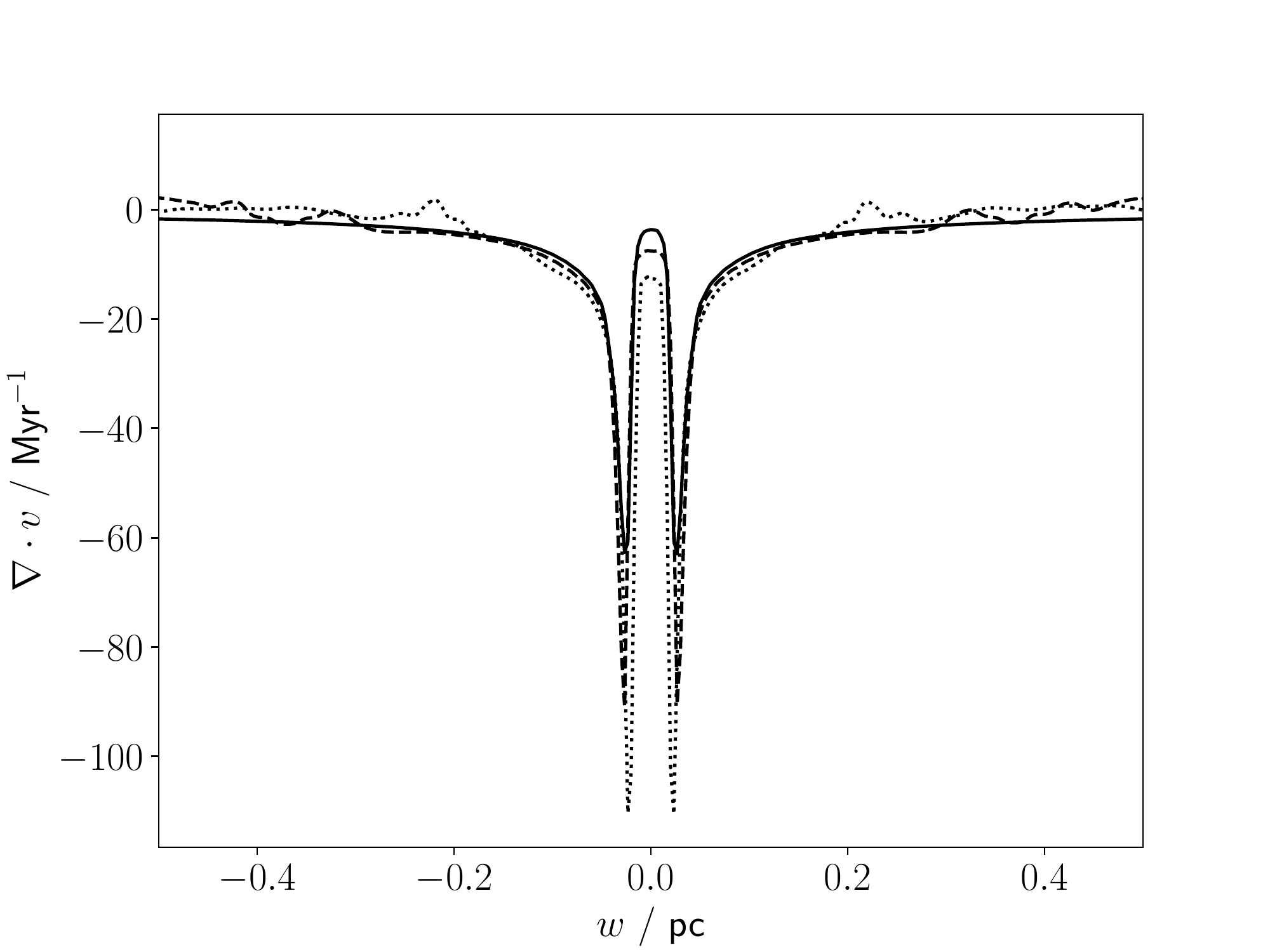}}\\
\caption{Velocity divergence profiles corresponding to the volume-density profiles in Figure \ref{FIG:VolDensProfs}.}
\label{FIG:VelDivProfs}
\end{figure*}

\section{Model}\label{SEC:Model}

{We adopt the following simple initial conditions for the dynamical formation of a filament: a uniform-density cylinder of gas, symmetric about the $z$ axis, with a uniform inward radial velocity converging on the $z$ axis, and an isothermal equation of state. For this model we need only specify two {\it dimensionless} configuration parameters, the ratio, ${\cal G}$,  of the line-density of the filament to the critical value, and the Mach Number, ${\cal M}$, of the initial inflow velocity. 

For the purpose of illustration, we can make the results {\it dimensional} by specifying just two physical variables, for which we choose the outer radius of the initial filament, $W\subO\!=\!1\,{\rm pc}$, and the isothermal sound speed in the filament, $a\subO\!=0.187\,{\rm km\,s^{-1}}$ (corresponding to molecular gas, with solar elemental composition, at $T\!\simeq\!10\,{\rm K}$).

If the initial density in the filament is $\rho\subO$, then the line-density of the filament is $\mu\!=\!\pi W\subO^2\rho\subO$. The critical line-density {for stability against gravitational collapse} is $\mu_{_{\rm CRIT}}\!=\!2a\subO^2/G$ \citep{StodolkiewiczJ1963,ostriker1964}, and so
\begin{eqnarray}
{\cal G}&=&\frac{\pi G\rho\subO W\subO^2}{2a\subO^2}\,.
\end{eqnarray}
Hence the initial density in the filament is
\begin{eqnarray}
\rho(w\leq W\subO)&=&\rho\subO\;\,=\;\,\frac{2a\subO^2\;{\cal G}}{\pi GW\subO^2}\,.\hspace{0.3cm}
\end{eqnarray}
With our illustrative choice of $W\subO$ and $a\subO$,
\begin{eqnarray}
  \rho\subO&\rightarrow&5.18\,{\rm M_\odot\,pc^{-3}}\;{\cal G}\,,\\
  n_{_{\rm H_2:O}}&\rightarrow&73.0\,\rm{H_2\,cm^{-3}}\;{\cal G}\,.
\end{eqnarray}

If the initial inflow velocity is $v\subO$, then
\begin{eqnarray}
{\cal M}&=&v\subO/a\subO\,,
\end{eqnarray}
and hence the initial velocity field is
\begin{eqnarray}
{\boldsymbol v}(w)&=&-\,v\subO\,\hat{\boldsymbol e}_w\;\,=\;-\,a\subO\;{\cal M}\;\,\hat{\boldsymbol e}_w\,.\hspace{0.5cm}
\end{eqnarray}
where $\hat{\boldsymbol e}_w$ is a unit vector pointing radially away from the $z$ axis. With our illustrative choice of $a\subO$,
\begin{eqnarray}
v\subO&\rightarrow&0.187\,{\rm km\,s^{-1}}\;{\cal M}\,.
\end{eqnarray}

The gas in the filament has sound speed
\begin{eqnarray}
a(w\leq W\subO)&=&a\subO\,.
\end{eqnarray}

We must also specify three numerical parameters, but these are chosen so that they have minimal effect on the results. First, our model has a finite length, $L\subO$, rather than {being infinitely long.} $\,L\subO$ is chosen so that the central portions of the filament have time to evolve, well before they are overrun by the inward motion of the ends of the filament \citep{PonAetal2012b,ClarkeSWhitworthA2015}. This is achieved by setting $L\subO =5W\subO$. Second, the filament must be contained within a larger computational domain, for which we adopt a rectanguloid box with sides $\Delta x\subO$ $=$ $\Delta y\subO$ $=$ $3W\subO$, $\Delta z\subO$ $=$ $6W\subO$. Finally, the regions of the computational domain outside the filament must be filled with a rarefied, high-temperature `lagging' gas, so that in the initial state there is approximate pressure balance across the boundary of the filament. This requires the specification of a contrast parameter, ${\cal C}\subO$. The density in the lagging gas is then set to $\rho\subO/{\cal C}\subO$ and the sound speed to $a\subO{\cal C}\subO^{1/2}$. We adopt ${\cal C}\subO =10$, so with our illustrative choice of $W\subO$ and $a\subO$,
\begin{eqnarray}
\rho(w>W\subO)&\rightarrow&0.5185\,{\rm M_\odot\,pc^{-3}}\;{\cal G}\,,\\
a(w>W\subO)&\rightarrow&0.591\,{\rm km\,s^{-1}}\,.
\end{eqnarray}}

\begin{figure*}
\centering
\subfigure{\includegraphics[width=\columnwidth]{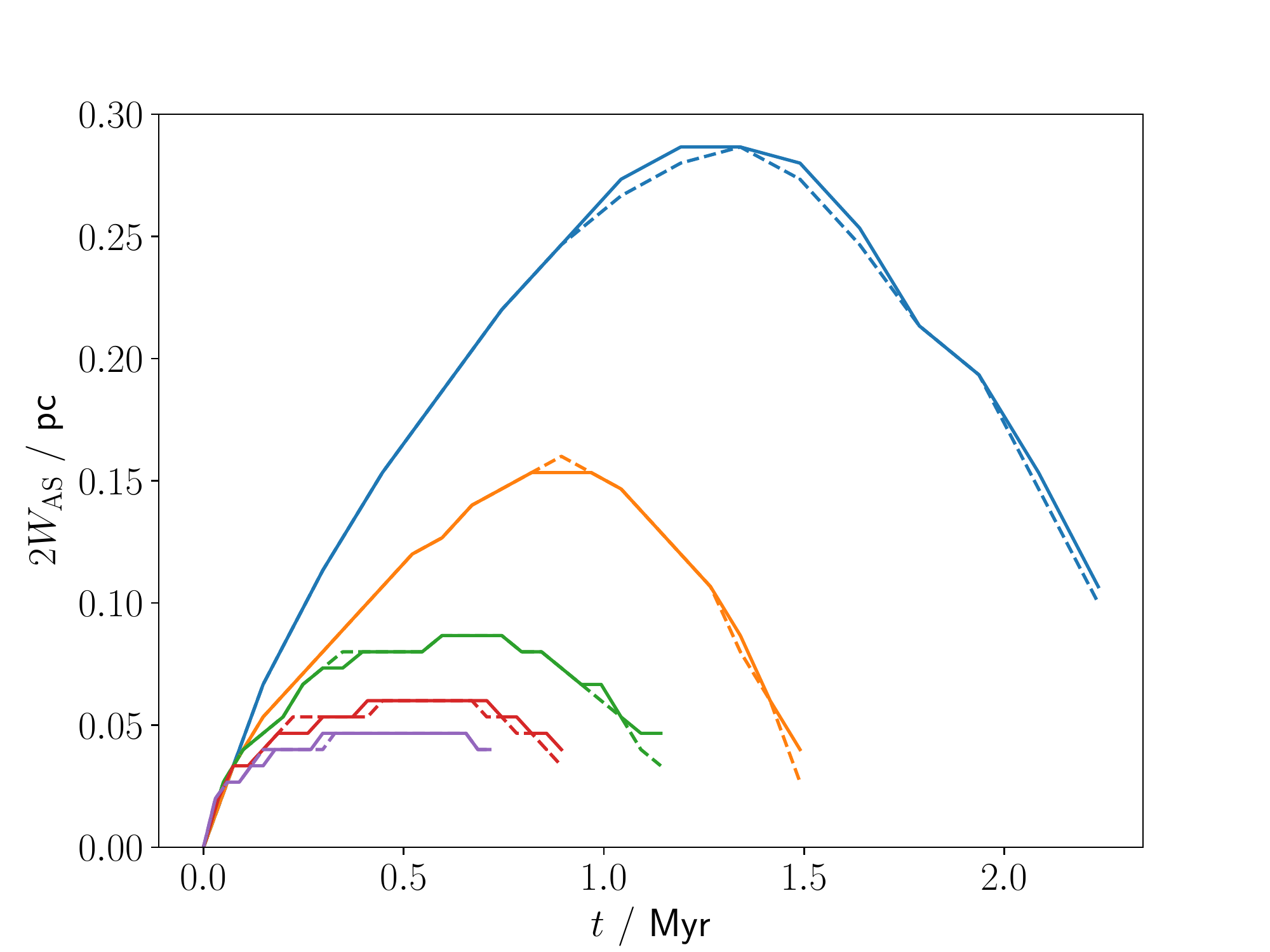}}\quad
\subfigure{\includegraphics[width=\columnwidth]{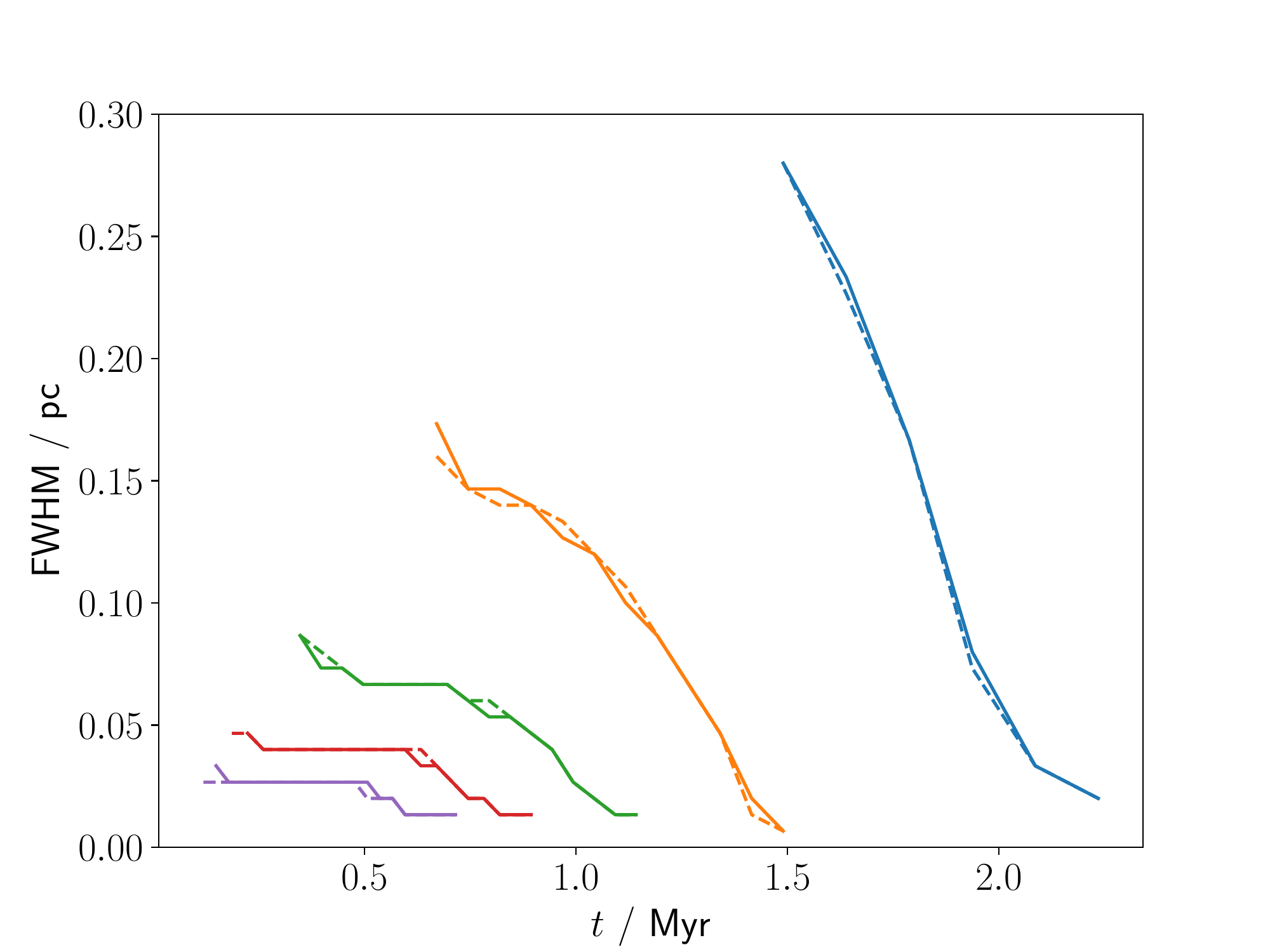}}
\caption{{\it Left panel:} evolution of $2\rdiv$ for filaments formed with different inflow Mach Numbers, ${\cal M}=1$ (blue), $2$ (orange), $3$ (green), $4$ (red), $5$ (purple). {\it Right panel:} evolution of $\fwhm$ for the same filaments as in the left panel. {Solid lines show properties of the averaged filament profiles, dashed lines those of the $z=0$ profiles.}}
\label{FIG:FWHM(t)aWSx2(t)}
\end{figure*}

For the dimensionless parameters, we investigate a fiducial value of ${\cal G} = 1.2$ (so that the filament is marginally supercritical) and ${\cal M} = 1-5$ (covering trans- to supersonic inflow velocities). We terminate the simulations when the maximum density reaches a value of $1.5 \times 10^7 \msun \pc^{-3}$ ($\sim 10^{-15} \gcc$). Beyond this time, $t\subFRAG$, the filament starts to fragment into prestellar cores \citep{ClarkeSetal2016}, so there is no longer a representative global filament profile to speak of. As the contracting filament does not remain entirely uniform in the $z$ direction (as a consequence of both imminent fragmentation, and longitudinal contraction due to its finite length), we construct radial profiles by averaging over the central region, $|z| \le 1 \pc$. {We obtain almost identical results if we instead consider the central $z=0$ profiles only (Figure \ref{FIG:FWHM(t)aWSx2(t)}).} We use the smoothed particle hydrodynamics code {\sc phantom} \citep{price2018}, with resolution parameter $\eta =1.2$, hence on average $\bar{\cal N}_{_{\rm NEIB}}\sim 56$ neighbours. We use $\sim 5\times 10^6$ particles, giving a particle mass of $\sim 2 \times 10^{-5} \msun$ and a notional mass resolution $\sim 0.001 \msun$ (i.e. one Jupiter mass).

\begin{figure}
\centering
\includegraphics[width=\columnwidth]{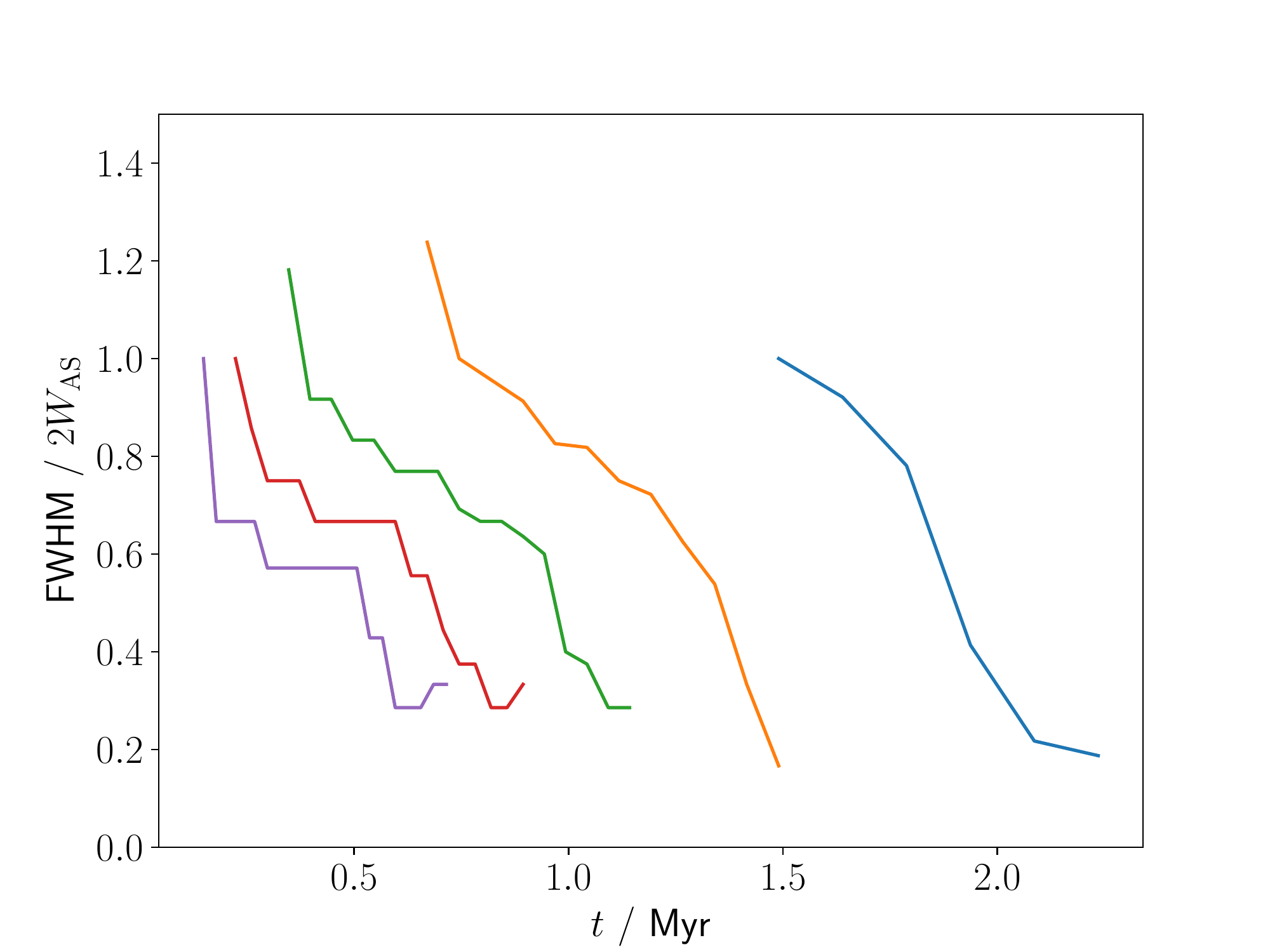}
\caption{Evolution of the ratio of $\fwhm$ to $2\rdiv$, for filaments formed with different inflow Mach Numbers, ${\cal M}=1$ (blue), $2$ (orange), $3$ (green), $4$ (red), $5$ (purple).}
\label{FIG:FWHMoWSx2(t)}
\end{figure}

\begin{figure}
\centering
\subfigure{\includegraphics[width=\columnwidth]{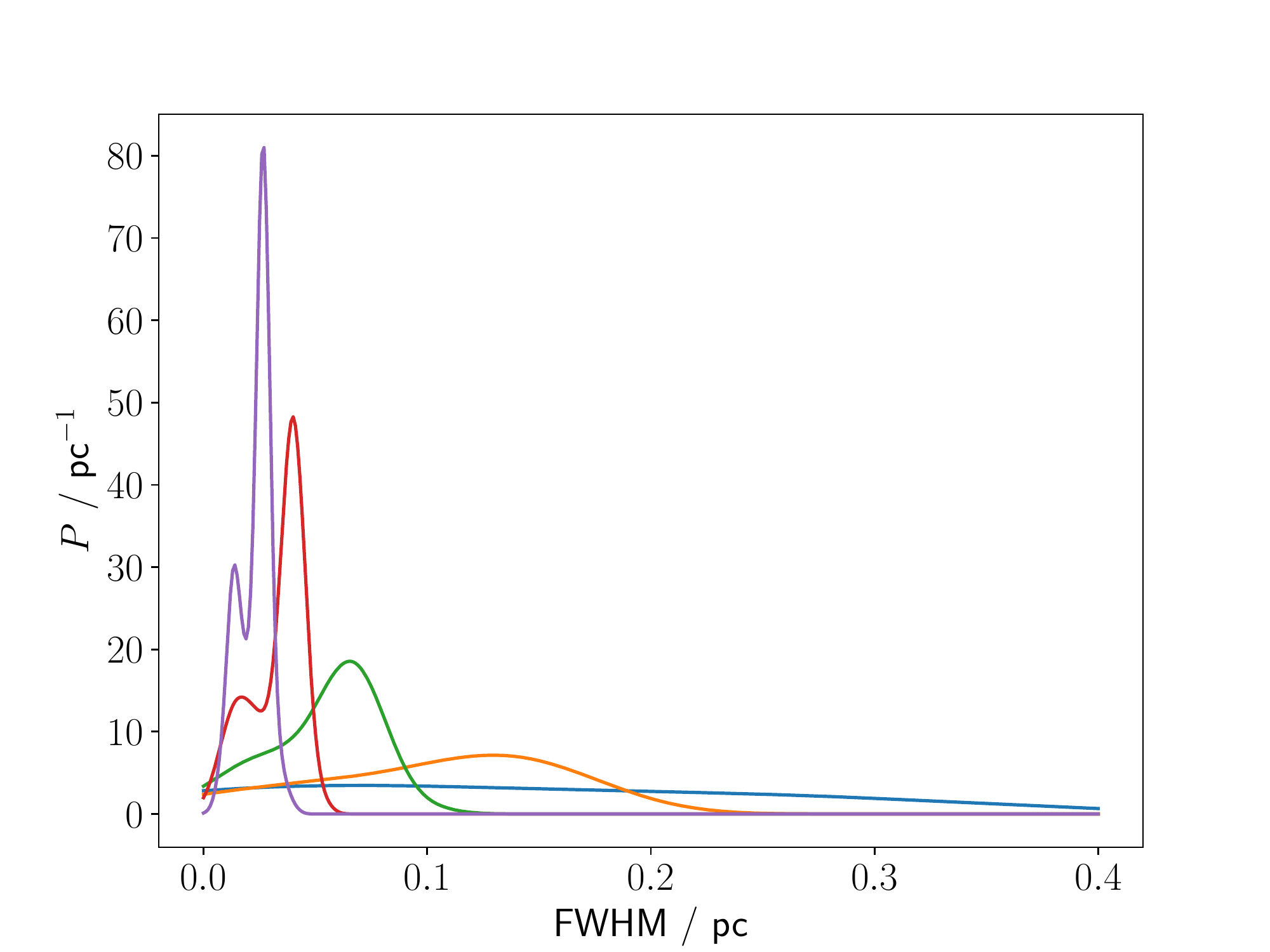}}\quad
\caption{The PDF for filament $\fwhm$s formed with different inflow Mach Numbers; ${\cal M}=1$ (blue), $2$ (orange), $3$ (green), $4$ (red), $5$ (purple).}
\label{FIG:Probs}
\end{figure}

\begin{figure*}
  \centering
  \subfigure{\includegraphics[width=0.66\columnwidth]{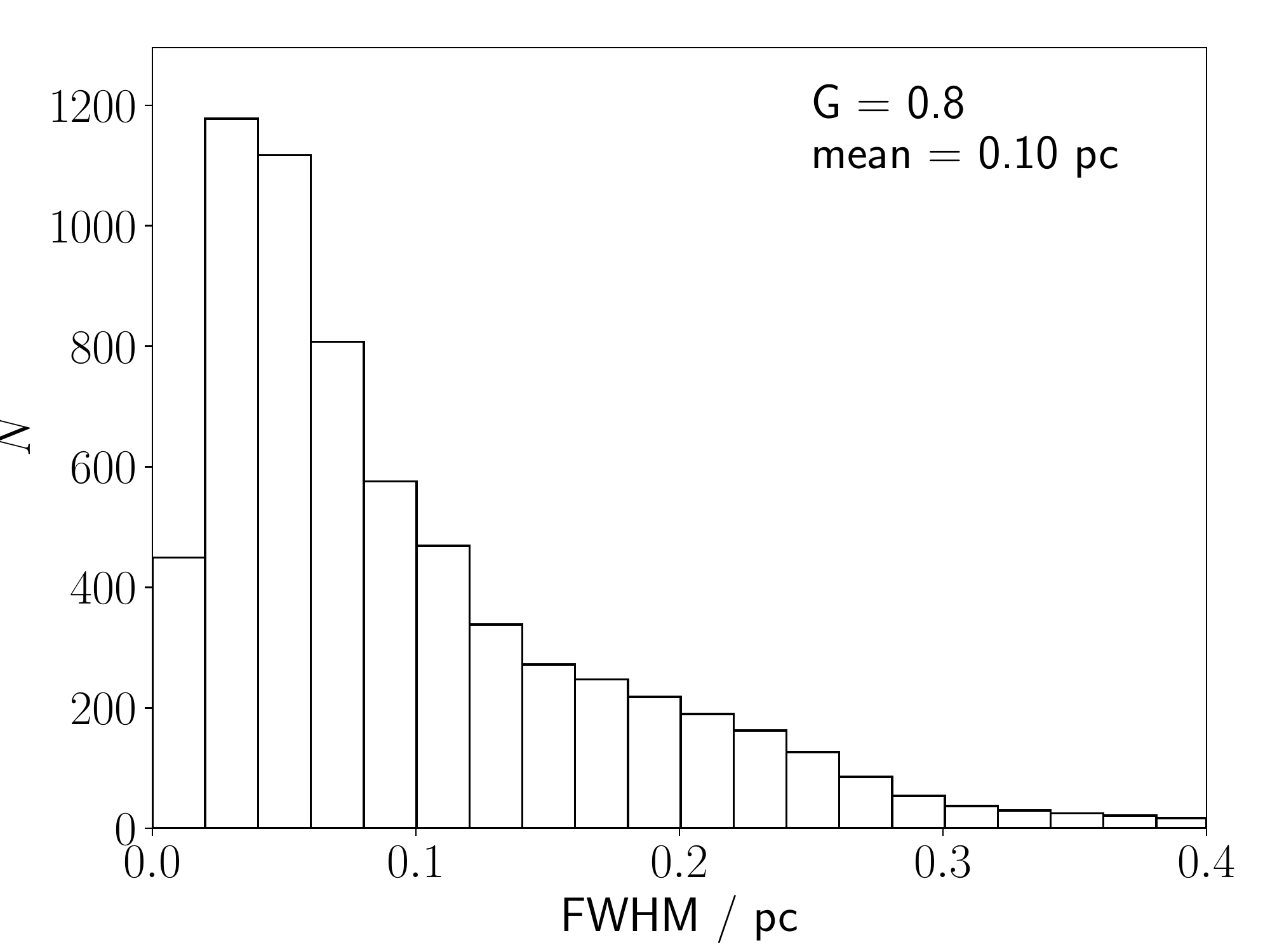}}\quad
  \subfigure{\includegraphics[width=0.66\columnwidth]{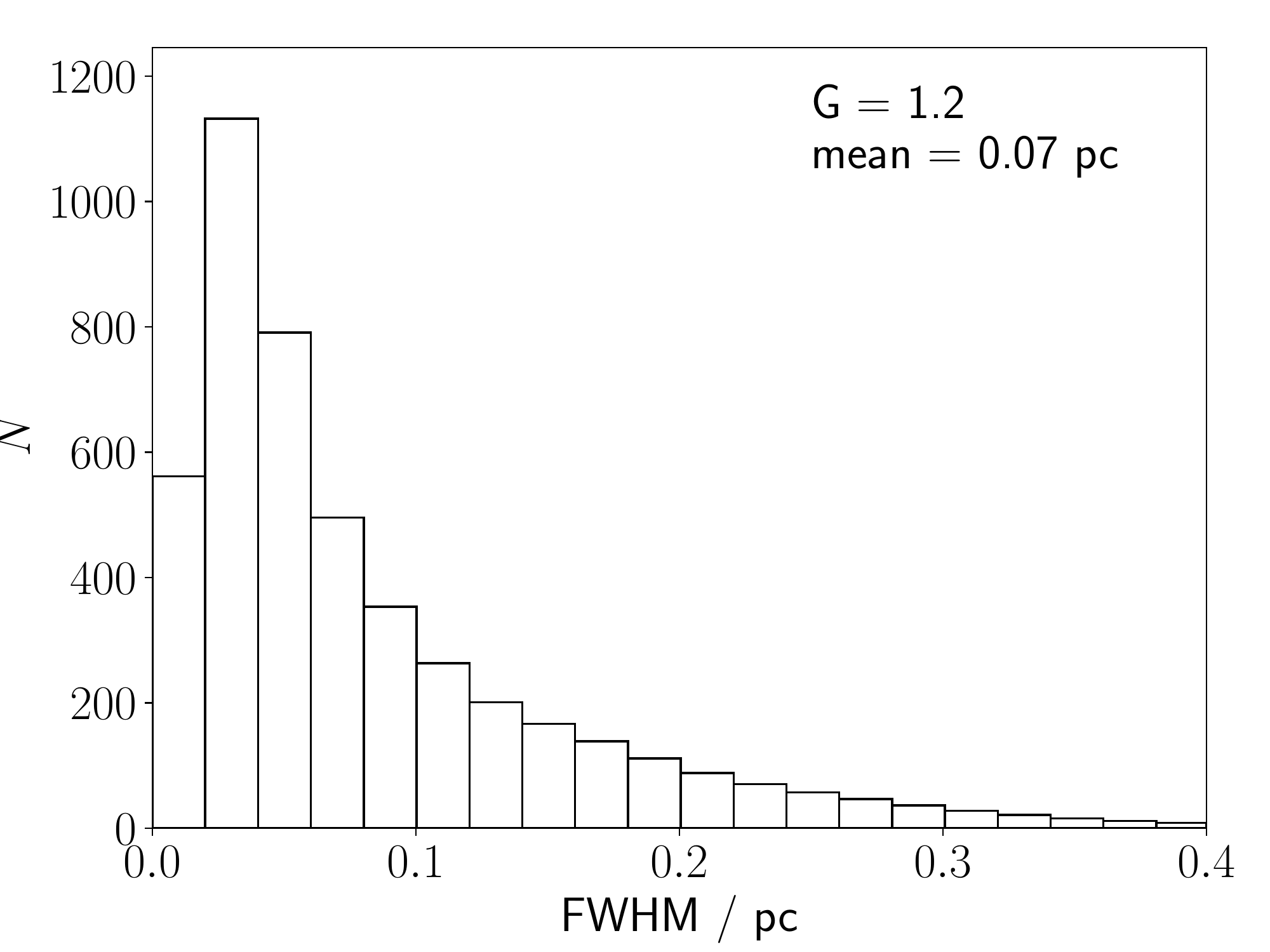}}\quad
  \subfigure{\includegraphics[width=0.66\columnwidth]{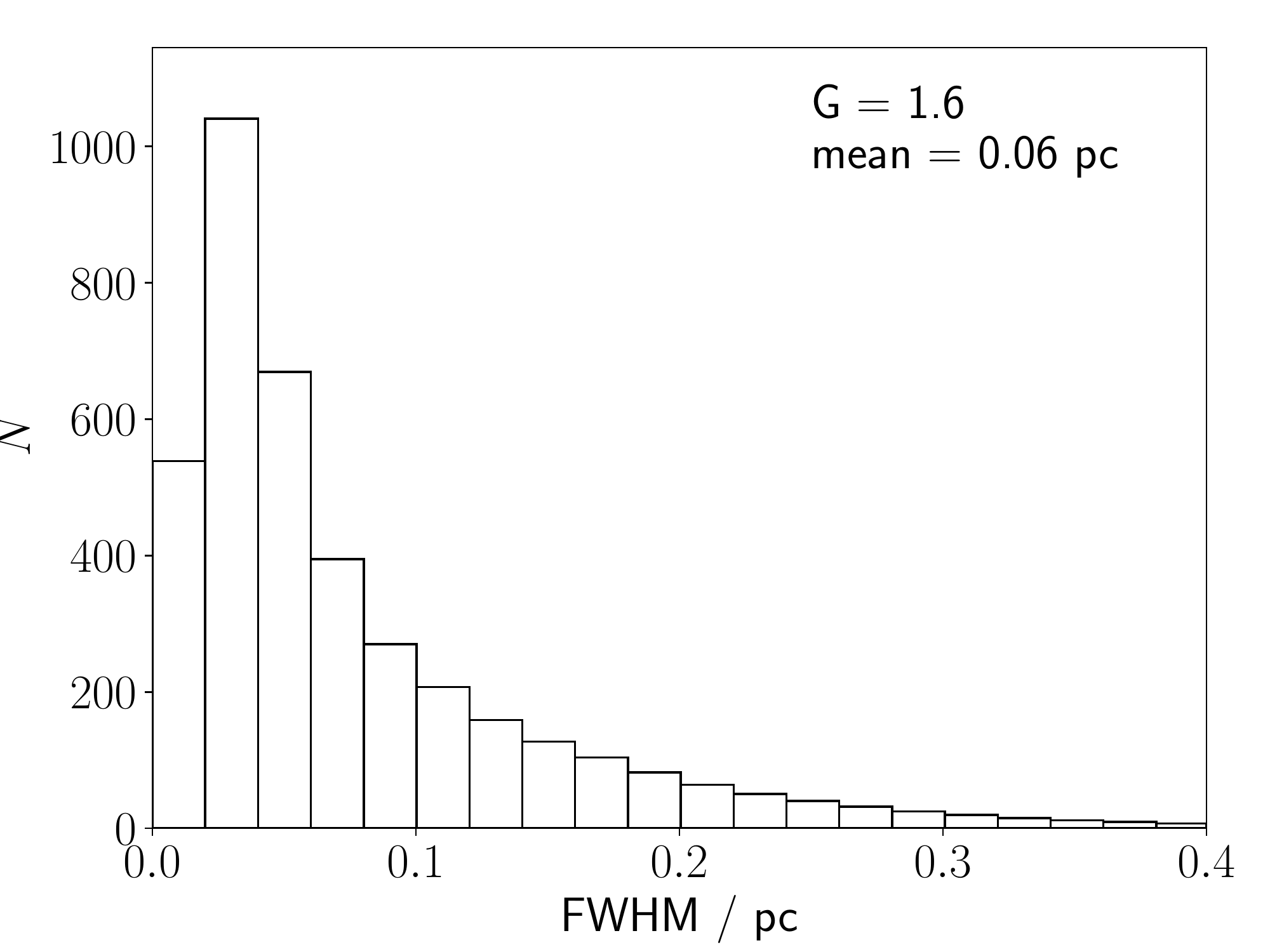}}\quad
  \caption{Histograms of filament $\fwhm$s for a uniform distribution of Mach Number $1 \le {\cal M} \le 5$, and ${\cal G} = 0.8$ (left), $1.2$ (middle) and $1.6$ (right).}
  \label{fig:histograms}
\end{figure*}

\begin{figure*}
  \centering
  \subfigure{\includegraphics[width=0.66\columnwidth]{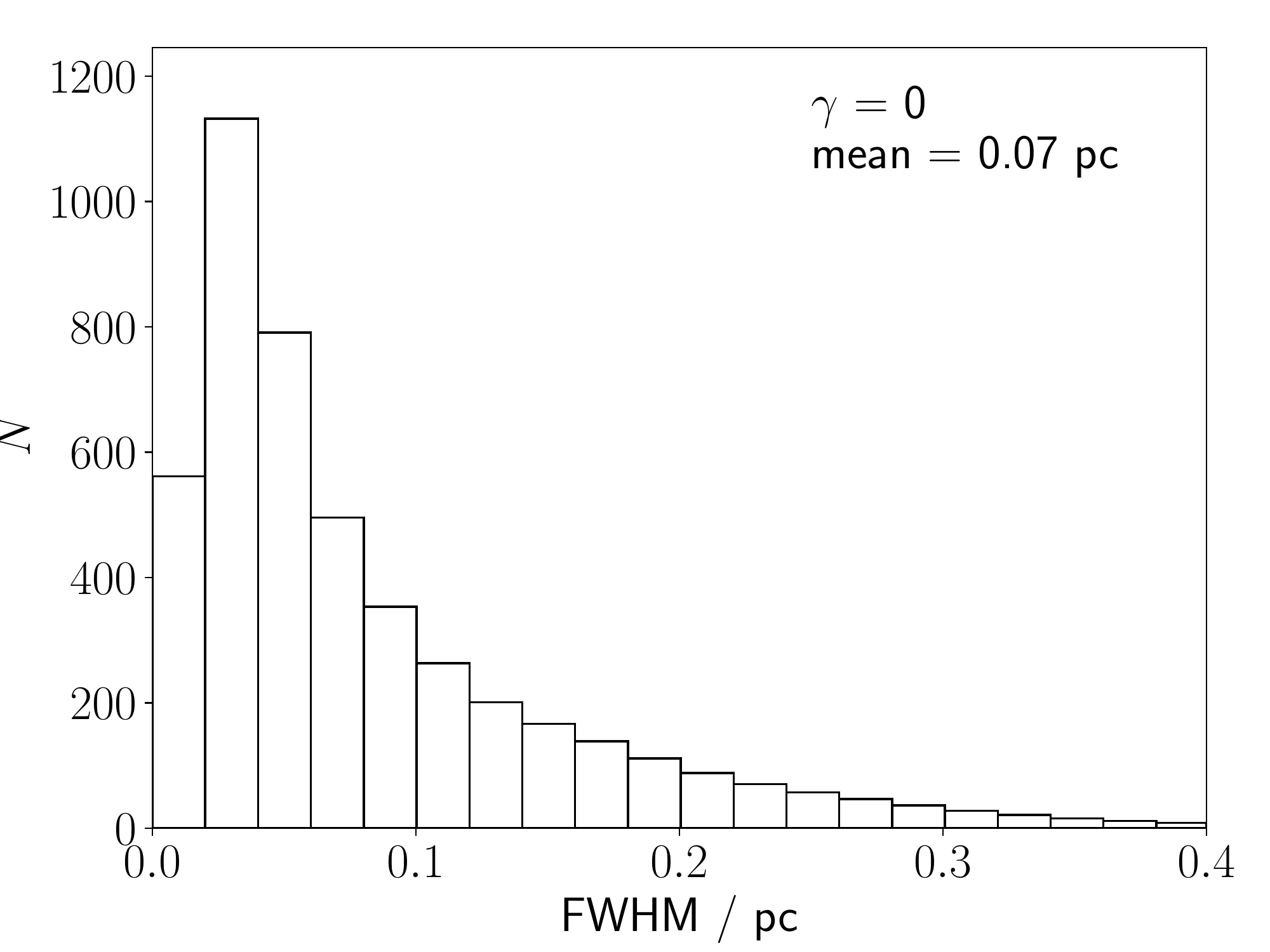}}\quad
  \subfigure{\includegraphics[width=0.66\columnwidth]{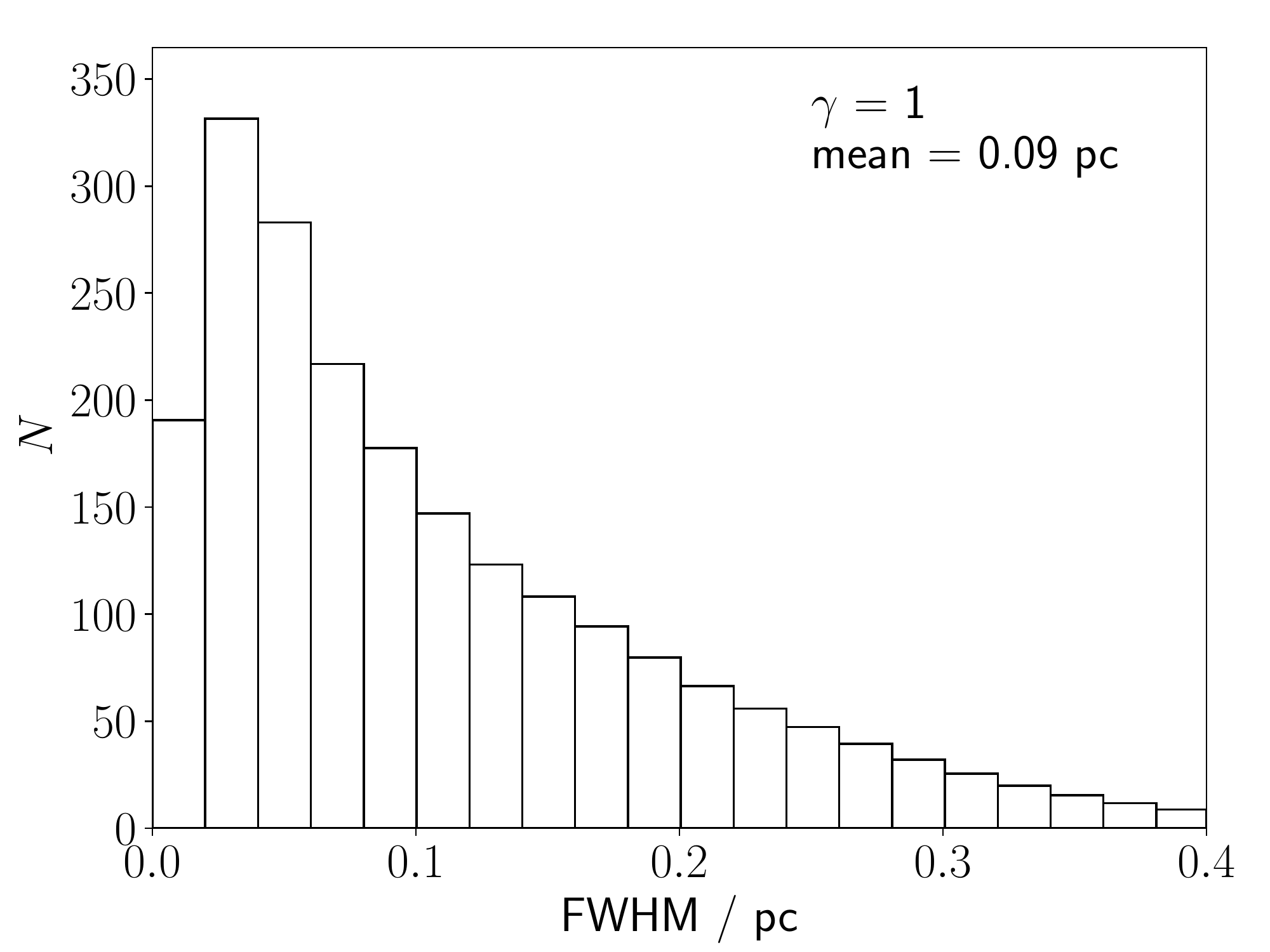}}\quad
  \subfigure{\includegraphics[width=0.66\columnwidth]{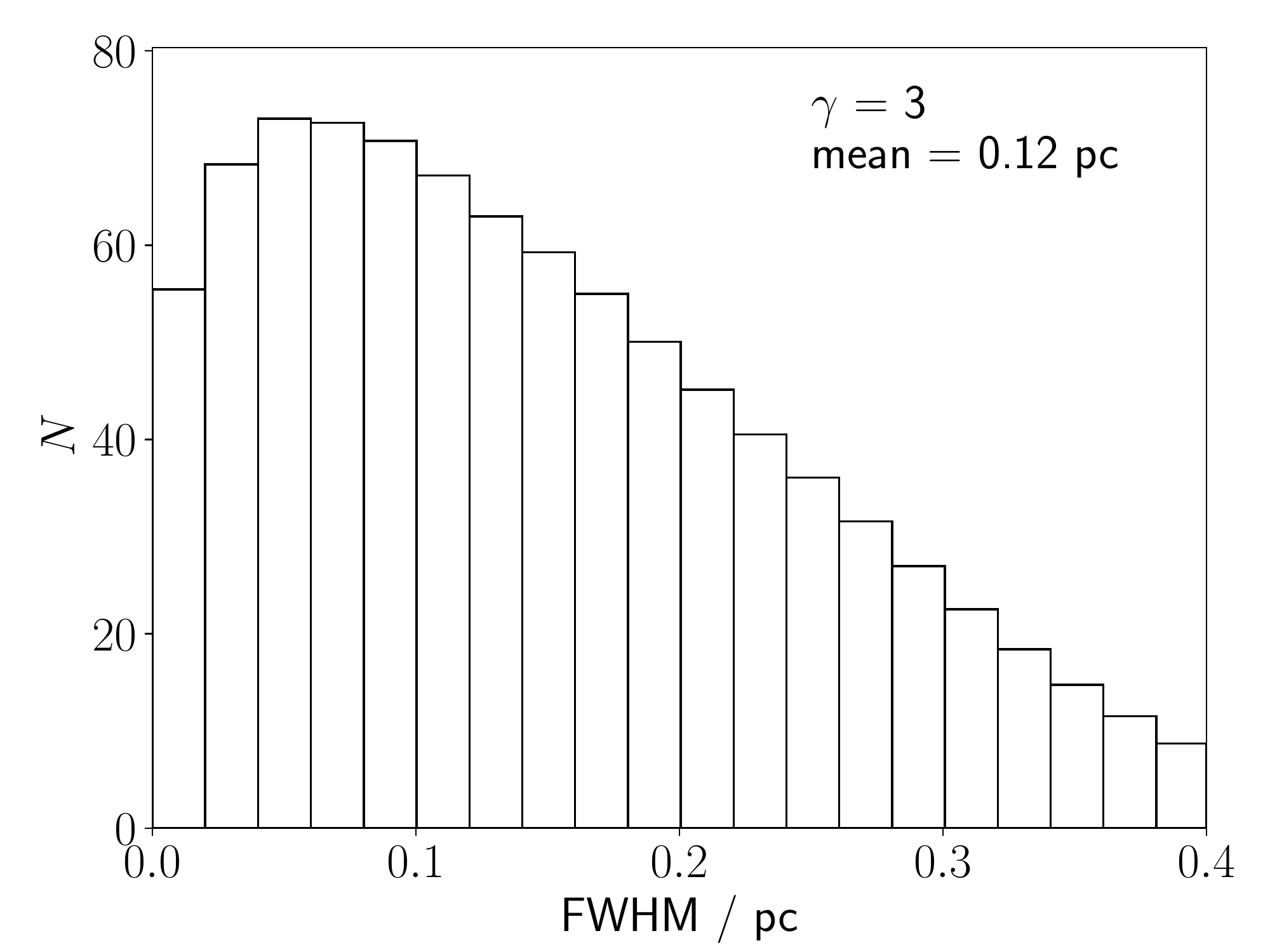}}\quad
  \caption{Histograms of filament $\fwhm$s for a distribution of Mach Number $\frac{dP}{d{\cal M}} \propto {\cal M}^{-\gamma}$ between $1 \le {\cal M} \le 5$, with $\gamma = 0$ (left), 1 (centre) and 3 (fight).}
  \label{fig:gamma}
\end{figure*}

\begin{figure}
\centering
\subfigure{\includegraphics[width=\columnwidth]{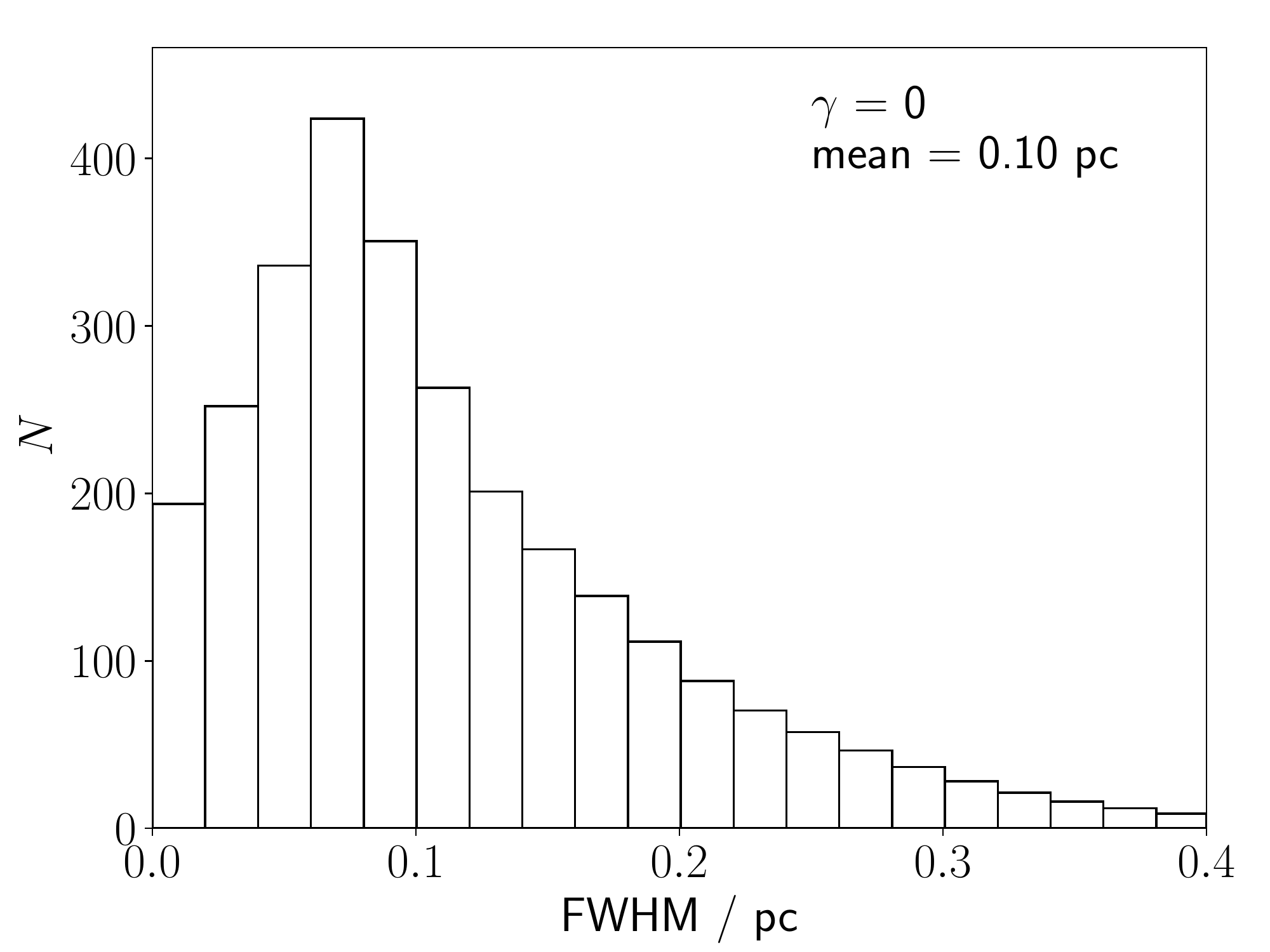}}\quad
\caption{Histogram of filament $\fwhm$s for a uniform distribution of Mach Number $1 \le {\cal M} \le 3$ and ${\cal G} = 1.2$.}
\label{fig:max3}
\end{figure}

\begin{figure*}
\centering
\subfigure{\includegraphics[width=\columnwidth]{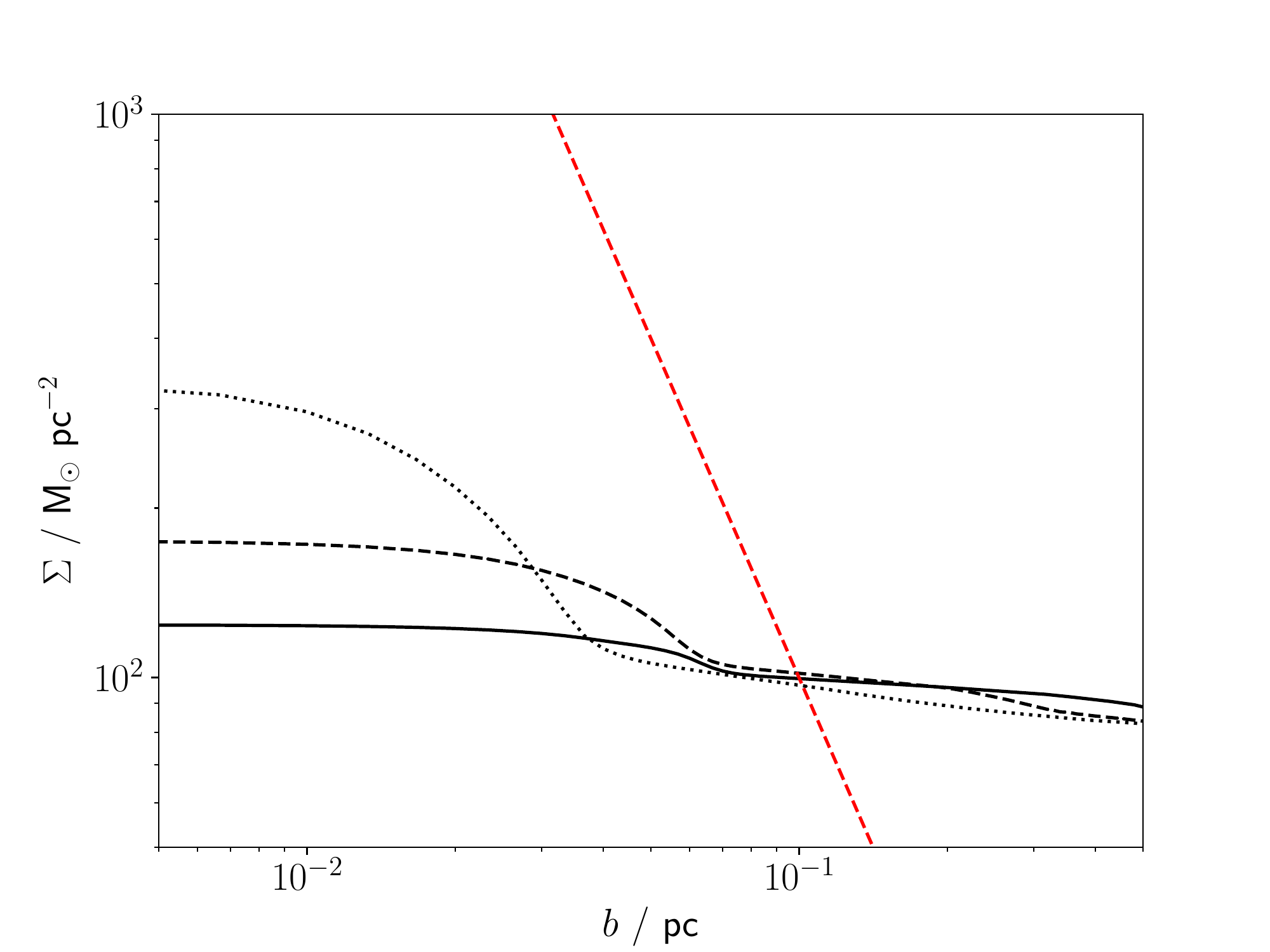}}\quad
\subfigure{\includegraphics[width=\columnwidth]{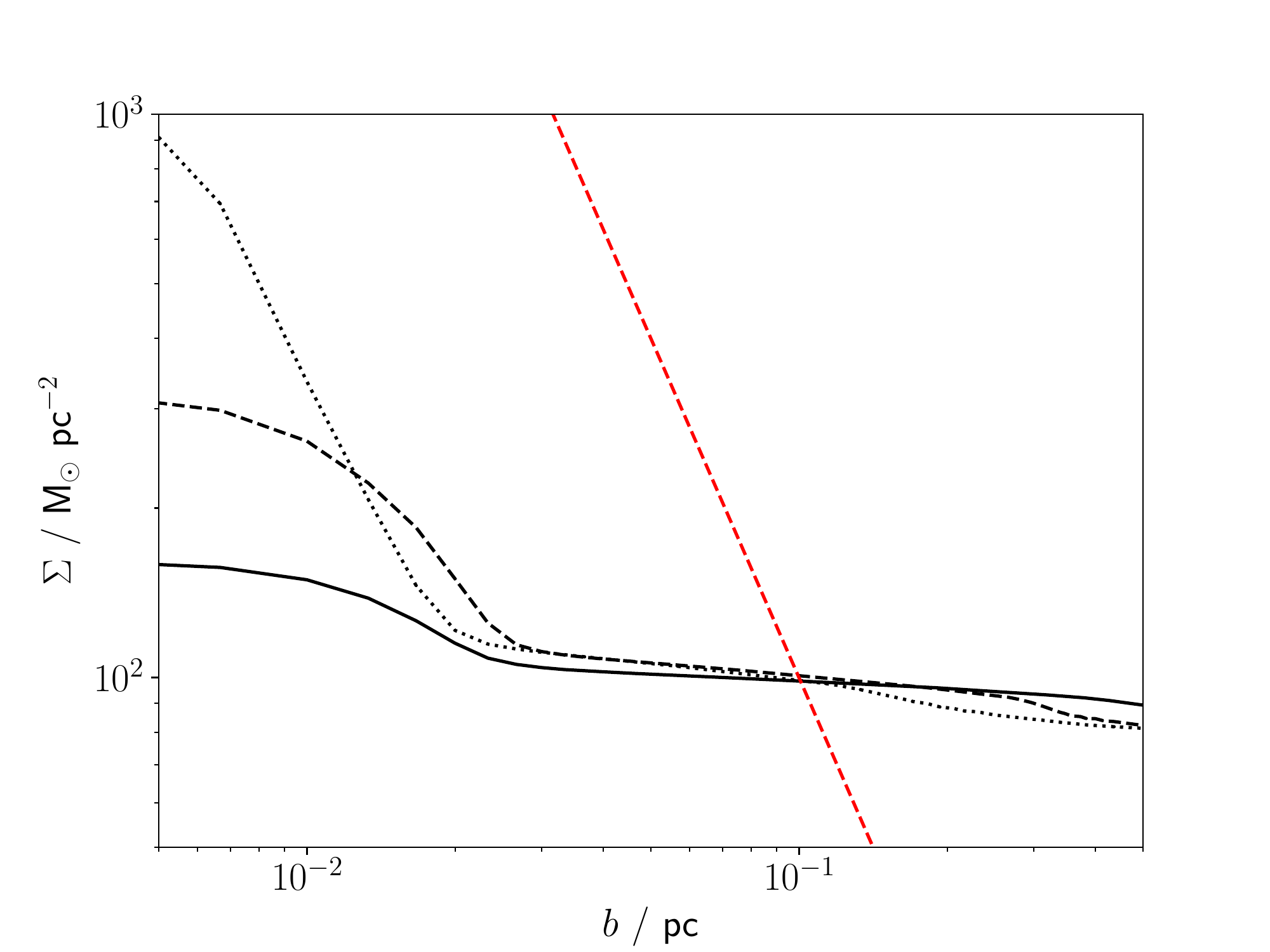}}\\
\caption{As Figure \ref{FIG:SurfDensProfs}, but with a logarithmic scale for the abscissa. The red dashed lines indicate a $\Sigma \propto b^{-2}$ profile.}
\label{fig:plum}
\end{figure*}

\begin{figure}
\centering
\includegraphics[width=\columnwidth]{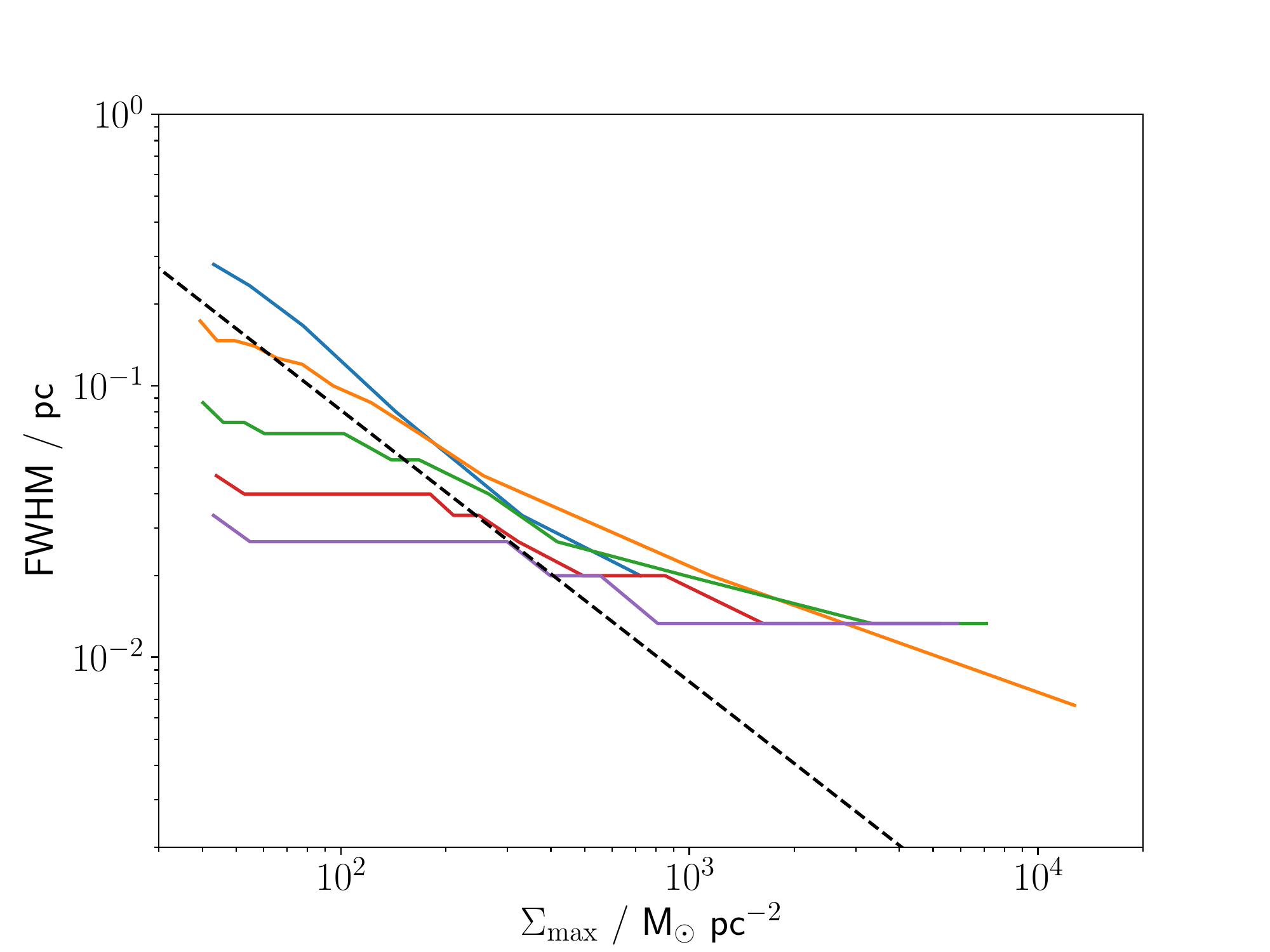}
\caption{{Filament $\fwhm$ versus central surface density, for filaments formed with different inflow Mach Numbers, ${\cal M}=1$ (blue), $2$ (orange), $3$ (green), $4$ (red), $5$ (purple). The black dashed line indicates the central Jeans length, $\lambda_{\rm J} = a\subO^2/(G\Sigma)$, as a function of surface density.}}
\label{fig:widthsigma}
\end{figure}

\section{Results}\label{SEC:Results}

\subsection{Filament profiles}

Figure \ref{FIG:VolDensProfs} shows volume-density profiles for the models with ${\cal M}=2$ (left panel) and ${\cal M}=4$ (right panel) at various times, up to and including the endpoints of the simulations (i.e. $t\subFRAG$). Figure \ref{FIG:SurfDensProfs} shows the corresponding surface-density profiles. Figures \ref{FIG:VelProfs} and \ref{FIG:VelDivProfs} show, respectively, the corresponding velocity and velocity-divergence profiles. 

For the volume-density profiles in Figure \ref{FIG:VolDensProfs}, we have added a uniform background volume-density equal to the volume-density in the initial filament, $\rho\subB=\rho\subO=6.21\,\msun\,\pc^{-3}$, in order to represent -- notionally -- the contribution from the larger cloud in which the filament is embedded. Similarly, for the surface-density profiles in Figure \ref{FIG:SurfDensProfs}, we have added a uniform background surface-density of $\Sigma\subB=80\,\msun\,\pc^{-2}$. The volume- and surface-density profiles (Figures \ref{FIG:VolDensProfs} and \ref{FIG:SurfDensProfs}) show qualitatively that the width of the filament does not vary much with time, especially for the higher Mach Number case.

\subsection{The accretion shock at the filament boundary}\label{SEC:AccShock}

The velocity profiles in Figure \ref{FIG:VelProfs} show that the inflow at large radii approximates to constant velocity. There is little {net inward acceleration of the inflow due to the combination of self-gravity (acting inward) and pressure gradient (acting outward)} until the gas hits the accretion shock, marking the boundary of the dense central filament. Most of the deceleration occurs at this accretion shock. There is some further deceleration inside the accretion shock, due to the pressure gradient that develops as the filament attempts to adjust towards hydrostatic equilibrium.

We posit that the accretion shock should be seen as the natural boundary of the filament. This separates objectively the more rarefied super- or trans-sonically inflowing material (not filament) from the much denser subsonic material (filament).

Because there is little acceleration at large radii, the density just outside the accretion shock at radius $\rdiv$ is approximately
\begin{eqnarray}\label{EQN:rhoPreS}
\rho\subpreS&\sim&\rho\subO[\rdiv/W\subO]^{-2}.
\end{eqnarray}
Eq. \ref{EQN:rhoPreS} simply reflects cylindrically symmetric convergence at constant velocity. The resolution in the gas just outside the accretion shock, as a fraction of the shock radius, is therefore
\begin{eqnarray}
\frac{4h\subpreS}{\rdiv}&\sim&4\eta\left[\frac{m\subSPH}{\rho\subO W\subO^2\rdiv}\right]^{1/3}\\
&\rightarrow&0.35\left[\frac{\rdiv}{0.01\,\pc}\right]^{-1/3}\!,\hspace{0.6cm}
\end{eqnarray}
where we have obtained the final expression by substituting the illustrative values of $\rho\subO$ and $W\subO$. It follows that the accretion shock is less well resolved when the filament is very narrow, but for all $\rdiv\ga 0.01\,\pc$ the code should be returning a meaningful estimate of the shock radius.

The velocity-divergence plots (Figure \ref{FIG:VelDivProfs}) allow us to locate the accretion shock quite accurately. We define the radius of the accretion shock, $\rdiv$, as the radius of the outermost divergence minimum with an absolute value at least five times the average.\footnote{This criterion is necessary because lower-${\cal M}$ models develop rarefaction waves propagating outward as incoming material `bounces' at the centre, and this results in additional minima in the divergence profiles inside that associated with the accretion shock.} The left panel of Figure \ref{FIG:FWHM(t)aWSx2(t)} shows the variation of the diameter of the filament, $2\rdiv$, with time, for filaments with different inflow Mach Numbers, ${\cal M}$. In the early stages, the filament's self-gravity is weak, and it is mainly contained by the ram-pressure of the inflow; consequently the diameter grows with time as the line-density increases. In the late stages, self-gravity dominates over ram-pressure, and the diameter decreases with time.

The diameter of the filament, $2\rdiv$, and the timescale for the filament to become gravitationally unstable against fragmentation, $t\subFRAG$, both decrease with increasing Mach Number of the inflow, due -- respectively -- to the greater ram-pressure (i.e. greater momentum-flux) and the greater mass-flux delivered to the accretion shock.

\subsection{The full-width at half-maximum}

The right panel of Figure \ref{FIG:FWHM(t)aWSx2(t)} shows the variation of the filament $\fwhm$ with time, for filaments with the same inflow Mach Numbers as those in the left panel. We only show values when the peak surface-density exceeds three times that of the original profile ($37 \msun \pc^{-2}$), {in order to ensure that the measured $\fwhm$ is that of the central filament, rather than the $\fwhm$ of the initial isolated cylinder as a whole}. {Once this threshold has been passed}, the variation of the $\fwhm$ echos closely the variation of the diameter, $2\rdiv$.

Figure \ref{FIG:FWHMoWSx2(t)} shows the ratio of the $\fwhm$ to $2\rdiv$, as a function of time. The $\fwhm$ is almost always smaller than $2\rdiv$, in other words the full-width at half maximum is almost always inside the accretion shock. As the line-density of the filament increases due to accretion, the filament becomes more self-gravitating and therefore more concentrated towards the spine. Consequently the ratio of the $\fwhm$ to $2\rdiv$ decreases with time, but {$\fwhm$ and $2\rdiv$ are within a factor of two of each other} for a significant fraction of the model evolution.

\subsection{The distribution of $\fwhm$s}\label{SEC:FWHMdistribution}

Figure \ref{FIG:Probs} shows the probability distribution function (PDF) of the $\fwhm$s for inflow Mach Numbers ${\cal M}$ $=$ 1, 2, 3, 4 and 5 -- i.e., for a given Mach Number, the probability of observing the filament when it has a given $\fwhm$, if the filament is observed at a random time between when it first exceeds our surface-density threshold and when it becomes unstable against fragmentation ($t\subFRAG$). These PDFs are generated from the discrete values of $\fwhm$ obtained from the simulations, smoothed using the procedure described in Appendix \ref{APP:Smoothing}. The higher the inflow Mach Number, the smaller the median $\fwhm$ and the narrower the PDF about this median value, due to the greater ram-pressure of the inflow.

Using the $\fwhm$ PDFs, we generate a large synthetic sample of filament $\fwhm$s, assuming a uniform distribution of Mach Numbers over the interval $1 \le {\cal M} \le 5$, and accounting for the shorter duration of higher-${\cal M}$ models and thus the lower probability of such a model being observed. This is directly comparable to the observational data from \citet{panopoulou2017}, assuming that real molecular clouds also contain filaments formed from flows of different ${\cal M}$ and at a range of evolutionary stages. Figure \ref{fig:histograms} shows $\fwhm$ histograms for filaments with ${\cal G} = 0.8$, $1.2$ and $1.6$. The overall shape of the $\fwhm$ distribution is not significantly affected by the gravitational {stability} of the filaments, and is very similar to that found by \citet{panopoulou2017}, with a peak at small $\fwhm$ and a gradual decline in the number of filaments with larger $\fwhm$. The formation of filaments by converging flows thus naturally produces a peaked distribution of filament widths, as observed.

The peak $\fwhm$ in Figure \ref{fig:histograms} is at $\sim 0.03 \pc$, rather than the $0.1 \pc$ observed. In Appendix \ref{APP:MachPDF}, we show that rather than a uniform distribution of Mach Number, there is likely to be a preference for low-${\cal M}$ flows in molecular clouds. Specifically, for a turbulent medium obeying standard scaling laws, we demonstrate (a) that the structures produced by converging flows tend to be filamentary, and (b) that the probability distribution of the Mach Numbers for filament-forming flows takes the form
\begin{eqnarray}\label{EQN:gamma.1}
\frac{dP}{d{\cal M}}&\propto&{\cal M}^{-\gamma},
\end{eqnarray}
with $\gamma$ in the range $0\lesssim\gamma\lesssim 3$. Figure \ref{fig:gamma} shows the impact of an increasingly bottom-heavy distribution of ${\cal M}$ on the $\fwhm$ distribution. While higher values of $\gamma$ do shift the peak $\fwhm$ to values closer to {those} observed, this comes at the expense of significantly more filaments with $\fwhm \gtrsim 0.2 \pc$, which are almost completely absent from the \citet{panopoulou2017} data. Obtaining a peak $\fwhm$ of $\sim 0.1 \pc$ from our model, while remaining consistent with other observational constraints, appears to require limiting the range of ${\cal M}$ to values $\lesssim 3$, as shown in Figure \ref{fig:max3} for $\gamma = 0$. Higher-velocity inflows produce strongly-peaked $\fwhm$ PDFs at values $\lesssim 0.05 \pc$ which are difficult to reconcile with the observed peak at higher values.

\section{Discussion}\label{SEC:Discussion}

\subsection{The peak of the $\fwhm$ PDF}

While our model of filament formation naturally explains the existence of a peak $\fwhm$ and the general shape of the distribution, the tendency to place this peak at too-low values $\fwhm$ unless high-${\cal M}$ flows are excluded is concerning. In Appendix \ref{APP:MachLimit}, we show that the \citet{LarsonRB1981} molecular cloud scaling relations result in a cut-off in the maximum Mach Number flow that can be present in a cloud of a given mass. For the \citet{PonAetal2012b} numerical values for the relations, the ${\cal M} \lesssim 3$ required by our model corresponds to molecular cloud masses $\lesssim 10^3 \msun$.

Hydrodynamical simulations of turbulent clouds produce both the correct average filament width \citep{federrath2016} and similar distributions of unaveraged widths \citep{priestley2020} to those observed, despite the initial presence of Mach 5 (and higher) motions. Additionally, the distribution of the velocity difference across filaments in \citet{priestley2020}, which corresponds approximately to $2v\subO$, drops off sharply above $1.0 \kms$, i.e. ${\cal M} \sim 3$ for sound speed $a\subO = 0.187 \kms$. This suggests that, at least in simulations, high-${\cal M}$ flows are not involved in forming the filamentary structures observed.

Filaments may also include dynamically-important magnetic fields, and we will explore this possibility in detail in a subsequent paper. If the field is initially uniform, $\boldsymbol{B}\subO$, it is likely to have two effects on the filament width. The component parallel to the filament spine, $\boldsymbol{B}_\parallel =[\boldsymbol{B}\subO\!\!\cdot\!\hat{\boldsymbol e}_z]\,\hat{\boldsymbol e}_z$, will give the filament extra support against collapse, resulting in a broader surface-density profile than the non-magnetised case. If there is also a component perpendicular to the filament spine, $\boldsymbol{B}_{\!\perp} =\boldsymbol{B}\subO -\boldsymbol{B}_\parallel$, it will provide additional support in directions perpendicular to $\boldsymbol{B}_{\!\perp}$ and $\hat{\boldsymbol e}_z$, thereby causing the filament to develop a flattened cross-section \citep[e.g.][]{KashiwagiRTomisakaK2021}; this in turn will increase the variance of $\fwhm$ values, since they will depend on the viewing angle, in addition to the Mach Number and evolutionary stage. We thus expect the introduction of a magnetic field to result in a broader distribution of $\fwhm$s peaked at higher values, which {might} bring our results in Figure \ref{fig:histograms} into closer agreement with the \citet{panopoulou2017} data. On the other hand, if the field has a toroidal component, this will probably act to compress the filament \citep[e.g.][]{FiegeJPudritzR2000a}.

{Finally, filaments may have additional support from turbulence, and -- all other things being equal -- this will increase their widths by supplying an extra internal pressure. Indeed, observed filaments are typically observed to have trans-sonic internal turbulence, i.e. non-thermal velocity dispersion between one and two times the isothermal sound speed, $c_{_{\rm S}}\lesssim\sigma_{_{\rm NT}}\lesssim 2c_{_{\rm S}}$ \citep[e.g.][]{HacarATafallaM2011, hacar2013, HacarAetal2017, hacar2018}. When this level of turbulence  is taken into account, a given inflow velocity, $v\subO$, translates into a Mach Number that is between $\sim\surd{2}$ and $\sim\surd{3}$ times smaller, and a filament width that is therefore between $\sim 2$ and $\sim 3$ times {larger}.}

\subsection{The Plummer exponent}

It is standard practice to fit the observed surface-density profiles of filaments with Plummer profiles, i.e.
\begin{eqnarray}
\Sigma(w)&=&\Sigma\subB\,+\,\Sigma\subO\left\{1\,+\,\left[\frac{w}{w\subO}\right]^2\right\}^{-[p-1]/2}.
\end{eqnarray} 
Here $\Sigma\subB$ is the background surface-density, $\Sigma\subB+\Sigma\subO$ is the surface-density on the spine of the filament, $w$ is the projected distance of the line of sight from the spine, $w\subO$ is the Plummer scale-length (related to the $\fwhm$ by $\fwhm =w\subO\{4^{1/[p-1]}-1\}^{1/2}$), and $p$ is the Plummer exponent. For an isothermal cylinder in equilibrium, $p=4$ \citep{StodolkiewiczJ1963,ostriker1964}, whereas observed filaments are typically best-fit with $p \sim 2$ \citep{arzoumanian2011,palmeirim2013,andre2014,arzoumanian2019}. This difference is often put down to effects such as magnetic fields or a non-isothermal equation of state \citep[e.g.][]{palmeirim2013}.

Figure \ref{fig:plum} shows the surface density profiles from Figure \ref{FIG:SurfDensProfs} on a log-log scale, along with a $\Sigma \propto b^{-2}$ profile (corresponding to $p = 3$; \citealt{casali1986}). As the central filament forms, the surface density profile in the outskirts gradually approaches $p = 3$ from lower values. {The averaging procedure used to estimate $p$ from observed filament profiles actually underestimates $p$ significantly \citep{whitworth2021}. Therefore the observed $p \sim 2$ may be explained, without recourse to additional physical processes, by a combination of filaments not being equilibrium objects and the averaging procedure used to estimate $p$.}

\subsection{The scaling of $\fwhm$ with $\Sigma$}

{\citet{arzoumanian2019} find no significant trend when comparing their observed filament widths with the filament central column densities. They argue that this is inconsistent with isothermal, hydrostatic cylinders, which should have a width scaling approximately with the local Jeans length, $\lambda_{\rm J} = a\subO^2/(G\Sigma)$. Figure \ref{fig:widthsigma} shows that our model filaments do have a declining $\fwhm$-$\Sigma$ trend, with denser filaments tending to be narrower, although this trend is much shallower than the $\Sigma^{-1}$ dependence of $\lambda_{\rm J}$. However, the \citet{arzoumanian2019} data extend only up to central column densities of $\sim 2 \times 10^{22} \pcs$, corresponding to $\sim 400 \msun \pc^{-2}$. For surface densities below this value, the trend in our model $\fwhm$s is mild, and hardly noticeable for the higher-${\cal M}$ models. The lack of scaling of filament width with central density is understandable if filaments are not in hydrostatic equilibrium, but are in fact dynamically-forming structures.}

\section{Conclusions}\label{SEC:Conc}

We have shown, using hydrodynamical simulations, that a cylindrically-symmetric, converging flow produces a central filament with a boundary set by the location of the accretion shock onto the filament. Because this accretion shock has a width $0.03 \pc \lesssim 2 \rdiv \lesssim 0.3 \pc$ for moderately-supersonic flows, the full-width at half-maximum of the filament surface-density profile also falls within this range of values. For a sample of filaments seen at different evolutionary stages and formed by flows with a range of Mach Numbers, the $\fwhm$ distribution naturally has a prominent peak value, and a similar shape to the observed distribution in molecular clouds. In order to reproduce the observed peak at $\sim 0.1 \pc$, our model requires {that} the range of Mach Numbers is limited to ${\cal M} \lesssim 3$, although the inclusion of magnetic fields {and/or trans-sonic turbulence} is likely to relax this requirement.
  
\section*{Acknowledgements}
FDP and APW acknowledge the support of a Consolidated Grant (ST/K00926/1) from the UK Science and Technology Facilities Council (STFC).

\section*{Data Availability}
The data underlying this article will be shared on request. All software used is publicly available.

\bibliographystyle{mnras}
\bibliography{DynFilFormExp1}

\appendix

\section{Smoothed probability distributions}\label{APP:Smoothing}

{With our models typically having $\sim 20$ $\fwhm$ values each, the appearance of a histogram of values can be strongly affected by the number and width of bins chosen. We thus instead present our results as smoothed PDFs, with the value at any point determined by a weighted sum over the discrete filament $\fwhm$s. This requires the choice of one parameter, a smoothing parameter $Q$ determining the range of influence of each discrete value. As a compromise between resolution and uncertainty, we use $Q(N) = {\rm ceiling}\left(\sqrt{N+\epsilon}\right)$ with $\epsilon=0.1$.}

We first sort all $N$ $\fwhm$ values $V$ in ascending order, and calculate the smoothing length $\sigma$ for each. This is given by
  \begin{equation}
    \sigma(i) = \left( V(i+Q) - V(i-Q) \right) / 2 \, , 
  \end{equation}
  except for the first and last $Q$ values, for which
  \begin{equation}
    \sigma(i; i \le Q) = Q \left( V(i+Q) - V(1) \right) / \left( Q + i - 1 \right)
  \end{equation}
  and
  \begin{equation}
    \sigma(i; i > N-Q) = Q \left( V(N) - V(i-Q) \right) / \left( Q + N - i \right)
  \end{equation}
  respectively. To avoid issues when neighbouring points have nearly-identical values, we then modify the smoothing length to $\sigma(i) = \sqrt{\sigma(i)^2 + <\sigma>^2}$, where $<\sigma>$ is the mean value for all points.

With $\sigma$ determined, the PDF is given by
  \begin{equation}
    P(x) = \frac{1}{N} \ssum_i W\left( \frac{|x-V(i)|}{\sigma(i)} \right) / \sigma(i)
  \end{equation}
  where $W$ is the smoothing kernel given by
  \begin{equation}
    W(s) =
    \begin{cases}
      \frac{2}{3} - s^2 \left(1 - \frac{s}{2}\right), & \quad  s \le 1; \\
      \frac{1}{6} \left(2-s\right)^3, & \quad  1 < s \le 2; \\
        0, & \quad  s > 2.\\
    \end{cases}
  \end{equation}

\section{The formation of filaments, and other structures, by converging flows}\label{APP:MachPDF}

In this appendix we consider the distribution of Mach Numbers for the turbulent flows creating filaments.

There are two distinct potential explanations for the predominance of filamentary structures in dense interstellar gas, and therefore in star-formation regions. Firstly, given a static, dense layer of gas, the dominant fragmentation mode involves elongated perturbations that evolve into filaments, as shown on the basis of nonlinear perturbation analysis by \citet{MiyamaSetal1987a}, {and using numerical simulations by \citet{MiyamaSetal1987b} and \citet{ balfour2015}.} Secondly, even in regions where gravity is unimportant, shock-compression due to random isotropic turbulence produces filaments. This appendix is concerned with this second explanation. We demonstrate why the filamentary mode is ubiquitous, even in relatively low-density gas, and derive an expression for the differential probability, $dP/d{\cal M}$, that a randomly chosen element of gas is being compressed by a convergent flow at a given Mach Number, $\sim\!{\cal M}$.

\subsection{Scaling laws for interstellar turbulence}

We adopt a turbulent power spectrum of the form
\begin{eqnarray}
\frac{dP}{dk}&\propto&k^{-\beta}.
\end{eqnarray}
Typically $\beta\sim 2$. The density of modes in 3D ${\boldsymbol k}$-space then goes as $k^{-[\beta+2]}$ -- and hence, with $\beta =2$, as $k^{-4}$.

If we now switch to length-scale, $L$ as the independent variable, with $k=L^{-1}$, $dk/dL=-L^{-2}$, we have
\begin{eqnarray}
\frac{dP}{dL}&=&\frac{dP}{dk}\,\left|\!\frac{dk}{dL}\!\right|\;\,\propto\;\,L^{[\beta-2]}.
\end{eqnarray}

To obtain the probability that an element of gas has Mach Number ${\cal M}$, we invoke a scaling law of the form
\begin{eqnarray}
{\cal M}&\propto&\sigma\;\;\propto\;\;L^\alpha,
\end{eqnarray}
where $\sigma$ is the velocity dispersion, and the first proportionality presumes that the gas is isothermal and therefore has uniform sound speed. For the original scaling laws proposed by \citet{LarsonRB1981}, $\alpha\!\simeq\!0.38$. For the modified scaling laws proposed by \citet{PonAetal2012b}, $\alpha\!=\!1/2$, {as found by \citet{myers1983}}.

It follows that 
\begin{eqnarray}
\frac{dP}{d{\cal M}}&=&\frac{dP}{dL}\,\left|\!\frac{dL}{d{\cal M}}\!\right|\;\;\propto\;\;{\cal M}^{[\beta -1-\alpha]/\alpha}.
\end{eqnarray}

We also define the timescale between compression events for such a gas element,
\begin{eqnarray}
t({\cal M})&\propto&\frac{L({\cal M})}{\sigma({\cal M})}\;\;\propto\;\;{\cal M}^{[1-\alpha]/\alpha}
\end{eqnarray}

\subsection{Shock-compressed sheets}

We define a `shock-compressed sheet' as a sheet whose containment is initially dominated on both sides by the ram-pressure of material accreting onto the sheet; only later, if at all, does containment become dominated by self-gravity. In the absence of stellar feedback -- in particular expanding H{\sc ii} regions, stellar-wind bubbles and supernova remnants -- this would appear to be the only way in which sheets can form in the interstellar medium.

A single converging mode is required to form a shock-compressed sheet, and therefore the frequency with which an element of the interstellar medium becomes part of a shock-compressed sheet is given by
\begin{eqnarray}
\left.\frac{d{\cal F}}{d{\cal M}}\right|_{_{\rm SHEET}}&\propto&\frac{1}{t({\cal M})}\,\frac{dP}{d{\cal M}}\;\;\propto\;\;{\cal M}^{[\beta -2]/\alpha}.\hspace{0.5cm}
\end{eqnarray}

However, an isolated extended sheet with an isothermal equation of state does not collapse, no matter how fast it is converging or how high the external pressure. In fact, this is true for any polytropic equation of state, provided the polytropic exponent is positive, i.e. $\left.d\ln(P)/d\ln(\rho)\right|_{_{\rm LAGRANGIAN}}>0$. Consequently shock-compressed sheets are necessarily transient (they form and then bounce) unless they are sufficiently massive and there is a sufficient external pressure (in which case they relax towards hydrostatic equilibrium). Collapse, and hence star formation, can only occur in a shock-compressed sheet if it interferes constructively with another converging mode. 

\subsection{Shock-compressed filaments}

We define a `shock-compressed filament' as one in which containment of the embryonic filament is initially dominated by the ram-pressure of material accreting onto the filament; only later, if at all, does it become dominated by self-gravity. (This is not the only way to form a filament. As mentioned at the start of this appendix, filaments can also form by gravitational fragmentation of sheets, but in that case they are not `shock-compressed'.)

To form a shock-compressed filament requires two approximately orthogonal converging modes to interact more-or-less simultaneously, and hence to combine constructively. Consequently the frequency with which an element of the interstellar medium becomes part of a shock-compressed filament is given by
\begin{eqnarray}\label{EQN:dF/dM.1}
\left.\frac{d{\cal F}}{d{\cal M}}\right|_{_{\rm FILAMENT}}\!&\!\propto\!&\!\frac{1}{t^2({\cal M})}\,\frac{dP}{d{\cal M}}\;\;\propto\;\;{\cal M}^{[\beta -3+\alpha]/\alpha}.\hspace{0.5cm}
\end{eqnarray}
There is a geometric factor buried in the constant of proportionality, which reflects how approximately orthogonal the two converging modes (with independent wave-vectors $\boldsymbol{k}_{_1}$ and $\boldsymbol{k}_{_2}$) have to be to avoid the increased density and pressure delivered by the first converging mode being able to squeeze out sideways before the second converging mode arrives. This factor is presumed to be independent of the Mach Numbers of the converging modes.

\subsection{Shock-compressed cores}

Finally, we define a 'shock-compressed core' as one in which containment of the core is initially dominated by the ram-pressure of the material accreting onto the core; only later, if at all, does it become dominated by self-gravity (and hence become a prestellar core).

To form a shock-compressed core requires (at least) three approximately orthogonal converging modes to interact more-or-less simultaneously, and hence to combine constructively. Consequently the frequency with which an element of the interstellar medium becomes part of a shock-compressed core is given by
\begin{eqnarray}
\frac{d{\cal F}}{d{\cal M}}_{_{\rm CORE}}&\propto&\frac{1}{t^3({\cal M})}\,\frac{dP}{d{\cal M}}\;\;\propto\;\;{\cal M}^{[\beta -4+2\alpha]/\alpha}.\hspace{0.5cm}
\end{eqnarray}
Again, there is a geometric factor buried in the constant of proportionality, which reflects how approximately orthogonal to the first two converging modes the third converging mode has to be to avoid the increased pressure delivered by the first two converging modes being able to squeeze out sideways before the third converging mode arrives \citep{lomax2016}. Again we presume that this factor is independent of the Mach Numbers of the converging modes.

Since the formation of a shock-compressed core requires (at least) three approximately orthogonal converging modes to interact more-or-less simultaneously, the formation of an isolated shock-compressed core is rather unlikely, since this would require all three converging modes to have very narrow fronts.\footnote{By the `front' of a converging mode we mean the extent of the mode perpendicular to its wave-vector.} If two of the converging modes are broad, the core will {\it de facto} be embedded in a filament. If all three (or more) of the converging modes are broad, the core will {\it de facto} be at the confluence of three (or more) filaments.

\subsection{The distribution of Mach Numbers for cylindrical convergent flows}

From Eq. \ref{EQN:dF/dM.1}, the probability of a given piece of gas being currently involved in a convergent cylindrical flow can be written in the form
\begin{eqnarray}\label{EQN:dF/dM.2}
\left.\frac{d{\cal F}}{d{\cal M}}\right|_{_{\rm FILAMENT}}&\propto&{\cal M}^{-\gamma},
\end{eqnarray}
where
\begin{eqnarray}
\gamma&=&\frac{3-\alpha -\beta}{\alpha}.
\end{eqnarray}

If we adopt $\beta\!=\!2.0\pm0.5$, the \citet{LarsonRB1981} scaling laws give
\begin{eqnarray}
\gamma_{_{\rm LARSON}}&\simeq&1.63\pm1.32,
\end{eqnarray}
and the \citet{PonAetal2012b} scaling laws give
\begin{eqnarray}
\gamma_{_{\rm PON}}&\simeq&1.00\pm1.00.
\end{eqnarray}
Thus $\gamma$ could lie anywhere between $\sim\!0$ and $\sim\!3$, with the higher $\gamma$ values corresponding to the \citet{LarsonRB1981}. scaling laws. {This means that there is probably a preference for filaments to be formed by low-Mach Number flows.}

\section{The range of Mach Numbers in turbulent clouds}\label{APP:MachLimit}

In this Appendix, we consider constraints on the range of Mach Numbers for the turbulent flows creating filaments.

The scaling relations derived by \citet{LarsonRB1981} give, after some trivial algebra, the mean volume-density and the mean one-dimensional velocity dispersion\footnote{Note that Larson gives the three-dimensional velocity dispersion, which is larger by a factor $3^{1/2}$.} in a cloud of mass $M$:
\begin{eqnarray}
\bar{\rho}(M)&\sim&3.85\times 10^3\,\rm{M_{_\odot}\,pc^{-3}}\,[M/\rm{M}_{_\odot}]^{-0.58},\\
\bar{\sigma}(M)&\sim&0.635\,\rm{km\,s^{-1}}\,[M/\rm{M}_{_\odot}]^{0.20}.
\end{eqnarray}
In addition, \citet{LarsonRB1985} has suggested that the mean temperature in the interstellar medium can be approximated by a polytropic equation of state, $T\!\sim\!17\,\rm{K}\,[\rho/10^{-20}\,\rm{g\,cm^{-3}}]^{-0.28}$, and so the mean isothermal sound speed in a cloud of mass $M$ is
\begin{eqnarray}
\bar{a}(M)&\sim&0.170\,\rm{km\,s^{-1}}\,[M/\rm{M}_{_\odot}]^{0.081}.
\end{eqnarray}
It follows that the initial radius and line-density of the cylinder forming a filament should be scaled according to the cloud mass,
\begin{eqnarray}
W\subO(M)&\sim&\left[\frac{2{\cal G}}{\pi G\bar{\rho}(M)}\right]^{1/2}\,\bar{a}(M)\\
&\sim&0.059\,\pc\;{\cal G}^{1/2}\,[M/\rm{M}_{_\odot}]^{0.34},\\
\mu\subO(M)&\sim&\frac{2\bar{a}^{2}\!(M)}{G}\\
&\sim&13.5\,\rm{M_{_\odot}\,pc^{-1}}\,{\cal G}\,[M/\rm{M}_{_\odot}]^{0.16}.
\end{eqnarray}
Moreover, the inflow Mach Numbers should be capped at, or at least must decline rapidly above,
\begin{eqnarray}
{\cal M}\subMAX (M)&\sim&\frac{\bar{\sigma}(M)}{2^{1/2}\bar{a}(M)}\\
&\sim&2.64\,[M/\rm{M}_{_\odot}]^{0.12}.
\end{eqnarray}
Thus inflows with high Mach Numbers are only common in relatively massive clouds.

If we repeat this analysis using the scaling relations proposed by \citet{PonAetal2012b} and a very slightly cooler equation of state, $T\!\sim\!12.5\,\rm{K}\,[\rho/10^{-20}\,\rm{g\,cm^{-3}}]^{-1/4}$, we obtain
\begin{eqnarray}
\bar{\rho}(M)\!&\!\sim\!&\!3.17\times 10^3\,\rm{M_{_\odot}\,pc^{-3}}\,[M/\rm{M}_{_\odot}]^{-1/2},\hspace{0.5cm}\\
\bar{\sigma}(M)\!&\!\sim\!&\!0.148\,\rm{km\,s^{-1}}\,[M/\rm{M}_{_\odot}]^{1/4},\\
\bar{a}(M)\!&\!\sim\!&\!0.139\,\rm{km\,s^{-1}}\,[M/\rm{M}_{_\odot}]^{1/16},\\
W\subO(M)\!&\!\sim\!&\!0.030\,\pc\;{\cal G}^{1/2}\,[M/\rm{M}_{_\odot}]^{5/16},\\
\mu\subO(M)\!&\!\sim\!&\!9.1\,\rm{M_{_\odot}\,pc^{-1}}\,{\cal G}\,[M/\rm{M}_{_\odot}]^{1/8},\\
{\cal M}\subMAX (M)\!&\!\sim\!&\!0.752\,[M/\rm{M}_{_\odot}]^{3/16}.
\end{eqnarray}
Despite our adopting a slightly cooler equation of state, the \citet{PonAetal2012b} scaling relations leads to even lower ${\cal M}\subMAX$ values. Specifically, ${\cal M}\subMAX \gtrsim3$ requires $M\gtrsim 1.6\times 10^3\,\rm{M}_{_\odot}$, and ${\cal M}\subMAX \gtrsim5$ requires $M\gtrsim 2.4\times 10^4\,\rm{M}_{_\odot}$. {These results suggest that the upper limits on ${\cal M}$ invoked in Section \ref{SEC:FWHMdistribution} might be attributable to the relatively low masses of the star-forming clouds in which most filament widths have been measured.}

\bsp	
\label{lastpage}
\end{document}